\documentclass[sigconf,nonacm]{acmart}
\renewcommand\footnotetextcopyrightpermission[1]{}
\usepackage{booktabs}
\usepackage{tabularx}
\usepackage{array}
\usepackage{graphicx}
\usepackage{xcolor}

\DeclareUnicodeCharacter{2212}{\textminus}
\DeclareUnicodeCharacter{03B1}{\ensuremath{\alpha}}
\DeclareUnicodeCharacter{00D7}{\ensuremath{\times}}
\DeclareUnicodeCharacter{2265}{\ensuremath{\geq}}
\DeclareUnicodeCharacter{2264}{\ensuremath{\leq}}
\makeatletter\g@addto@macro{\UrlBreaks}{\do\-\do\/\do\a\do\b\do\c\do\d\do\e\do\f\do\g\do\h\do\i\do\j\do\k\do\l\do\m\do\n\do\o\do\p\do\q\do\r\do\s\do\t\do\u\do\v\do\w\do\x\do\y\do\z}\makeatother

\begin{document}

\title[Dependence, Imitation or Herding?]{Structural Dependence, Imitation or Herding? A Multi-Agent Simulation of Global Intermedia Agenda Setting}

\author{Mohammad Yousuf}
\orcid{0000-0003-3784-3346}
\affiliation{%
  \department{Department of Communication \& Journalism}
  \institution{University of New Mexico}
  \city{Albuquerque}
  \state{New Mexico}
  \country{USA}}

\author{Arif Md Tareque Habib}
\affiliation{%
  \department{Department of Communication \& Journalism}
  \institution{University of New Mexico}
  \city{Albuquerque}
  \state{New Mexico}
  \country{USA}}

\renewcommand{\shortauthors}{Yousuf and Habib}

\begin{abstract}
News outlets in poorer countries cover the world much as the leading outlets of rich countries do. Two explanations compete. One says they depend on Western news agencies for lack of resources. The other says they choose to copy prestigious outlets or whoever moves first. Observation cannot tell these apart. This study measured how closely 18 national outlets in Bangladesh, Kenya, South Africa, India, Malaysia and Nigeria followed four core outlets in how much attention they gave, which angle they took and when they published, on the wars in Ukraine and Gaza and two UN climate conferences in 2022–2024. It then ran a simulation in which language models acted as the outlets' editors and dependence, imitation and herding were switched on and off. Of nine tests, one held: outlets that relied more on wire copy published sooner after the core did. Tier and local stake showed no effect distinguishable from none. The herding test was supported and the imitation test was not. The simulation did not decide between the mechanisms, because the dependence rule, as fitted, reduced to publishing wire copy. Wire copy sets the clock of the periphery's foreign news. Whether it sets the agenda stays open.
\end{abstract}

\maketitle

\section{Introduction}

\subsection{Following without orders}

Studies of global news flow find that a small number of Western countries, defined theoretically as core countries, dominate global news coverage \cite{chang-2000-united-states} and shape how people elsewhere perceive the world \cite{guovargo2017-global-intermedia}. The imbalance persists online, where news outlets in poorer, peripheral countries follow that lead \cite{himelboim-2010-international-network}. Developing countries' newspapers have long depended on Western agencies for the bulk of their international news and reflected the news values of Western prestige dailies \cite{meyer-1989-global-news}, and recent South African online coverage of the war in Ukraine drew over 80 percent of its sources from Western newswires \cite{radebe-2022-features-media_abs}. Within one country, the international agendas of three U.S. evening newscasts correlated with that morning's \emph{New York Times} \cite{golan-2006-inter-media}. This study calls the pattern following and observes it at three levels, the attention an outlet gives an issue (salience), the angle it takes on it (framing) and when it covers it (timing). No wire agency or Western newspaper orders a newsroom in Dhaka, Lagos or Nairobi to follow. Why, then, does the periphery follow?

\subsection{Structure or choice}

Two families of answers compete, structure and choice. The structural answer holds that peripheral newsrooms lack the resources to gather foreign news and therefore depend on the news agencies of wealthy countries, above all the Associated Press (AP), Agence France-Presse (AFP) and Reuters. Pakistani newspapers hardly try to appoint correspondents and copy what the three agencies make available \cite{shaheen-2021-international-news}, and journalism practice in India is routinized around the same agencies \cite{mitra-2016-social-construction}. The answer from choice holds that newsrooms copy because they want to. Institutional theory holds that organizations facing uncertainty model themselves on others they see as more legitimate or successful \cite{dimaggio-1983-iron-cage_abs,sehl-2021-public-service}, which this study calls prestige imitation. An actor may also follow whoever moved first, without regard to its own information. Economists call this an informational cascade \cite{bikhchandani-1992-theory-fads_abs} and this study calls it herding. Argentine journalists monitor and imitate a shared pool of hard-news stories in that way \cite{boczkowski-2010-work-news}. The rival inside the structural camp is national interest, the pull of a country's own concerns on the foreign news it selects, which \citet{skurnik-1981-look-foreign} set against the dependence thesis.

Observational studies cannot separate these answers. Research on how one outlet's agenda shapes another's, known as intermedia agenda setting, mostly confirms a flow from elite to non-elite media \cite{su-2021-mapping-intermedia_abs}, and across countries a flow from high-power countries to the rest \cite{cui-2017-inter-media,guovargo2020-predictors-international}. Each finding is a correlation between agendas, and dependence, prestige imitation and herding all predict one. Tracing which outlet repeats a quote another used earlier can expose influence \cite{schlessinger-2025-exposing-obscured}, but not its mechanism. Separating the mechanisms needs a design in which one can be switched off while the others stay on. Agent-based models are computer models in which each actor, here each outlet, is a separate decision-maker that follows explicit rules, and such models of the news have been compared with observed coverage \cite{waldherr-2014-emergence-news}. Simulations whose agents are run by language models have varied one mechanism at a time on real news \cite{johansen-2025-imitating-agents}.

\subsection{This study}

This study tests the structural account against its rivals in two steps. First, it measures how closely 18 English-language national outlets in six countries followed four core outlets on three global issues from February 24, 2022, to December 31, 2024, tracking AP, AFP and Reuters as the source of wire copy, the ready-made stories agencies sell to newsrooms. The tiers come from world-system theory. \citet{wallerstein-1974-modern-world} described one world economy in which core areas exchange their manufactures for the raw materials of peripheral areas, with a semi-periphery between them. News research places countries in these tiers by the size of their economies and exports \cite{gunaratne-2001-prospects-limitations} and finds that core countries have a higher chance of being in the news \cite{chang-1998-countries-created}. On World Bank data for 2022 \cite{worldbank-2024-wdi}, Bangladesh, Kenya and South Africa fall in the periphery and India, Malaysia and Nigeria in the semi-periphery. The three issues differ in local stake, the degree to which an issue bears on a country's own interests. Stake is low for the war in Ukraine, which South African and Nigerian online media covered largely through Western wires and diaspora effects \cite{radebe-2022-features-media_abs,omoera-2023-sahara-reporters_abs}. It is mixed for the war in Gaza, which Hindi newspapers covered with India's own interests in view \cite{jangra-2024-hindi-newspapers_abs,baden-2016-convergent-news_abs}. It is high for the COP27 and COP28 climate summits, which drive climate coverage in India as elsewhere \cite{schfer-2013-what-drives_abs,gurwitt-2017-global-issue_abs}. Second, the study builds a multi-agent simulation in which language models act as the editors of the 18 outlets \cite{park-2023-generative-agents_abs}, switches dependence, prestige imitation and herding on and off, and scores each version on how well it reproduces the observed following. The simulation can show that a mechanism is sufficient to produce the pattern, not that newsrooms decide this way.

This study contributes to research on intermedia agenda setting, world-system theory and global news flow. Its novel feature is the use of large language model agents to simulate editorial decision-making in 18 newsrooms and to test, one mechanism at a time, why peripheral outlets follow the core. Studies of intermedia agenda setting show that agendas correlate across outlets and countries. Studies of news flow and the world-system explain that correlation by a country's structural position. Neither can tell dependence from prestige imitation or herding, because all three predict the same correlation. The methodological contribution is the combination of these theories with an agent-based simulation driven by language models. World-system theory supplies the tiers and the dependence mechanism. Institutional theory and cascade models supply its rivals. Intermedia agenda setting supplies the three levels of following, salience, framing and timing, against which each simulated mechanism is scored.

\section{Related Work}

\subsection{Dependence on the core's agencies}

Western agencies supply much of the foreign news the periphery prints, but why the periphery follows them is contested. Developing countries' newspapers depended on Western agencies for the bulk of their international news \cite{meyer-1989-global-news}, and in South African online coverage of the war in Ukraine over 80 percent of sources came from Western newswires \cite{radebe-2022-features-media_abs}. Countries in the core have a higher chance of being in the news than those below them \cite{chang-1998-countries-created,grasland-2019-international-news}, and media in high-power countries shape the global news agenda \cite{cui-2017-inter-media}, though a study of 11 countries concluded that the simple powerful-West reasoning may need to be reexamined \cite{du-2012-intermedia-agenda}. Structural position has been tested as a predictor of which countries are covered \cite{guovargo2020-predictors-international}. As the reason newsrooms follow the core, it has been assumed rather than tested against rival mechanisms.

The structural account starts from the world-system. \citet{wallerstein-1974-modern-world} divided one world economy into core, semi-periphery and periphery, and \citet{gunaratne-2001-prospects-limitations} ranked countries on that continuum by a weighted score of gross national product and exports, which turns position into a variable. The account's claim about newsrooms is that they rely on the core's agencies because they cannot afford to gather foreign news themselves. Sending correspondents abroad costs far more than an agency subscription \cite{mitra-2016-social-construction}, Rwandan newsrooms with limited finances depend on large foreign media \cite{bakina-2017-influence-foreign_abs}, and only a handful of news organizations, mainly Western, can afford international bureaus \cite{camaj-2010-media-framing}. Single-country studies document the reliance in both tiers. In \emph{The Star} of Malaysia and \emph{The New Vision} of Uganda, Western agencies played an overwhelmingly greater role than national ones \cite{adikata-1999-international-news}. All the stories about the West in two leading Nigerian newspapers came from Western agencies \cite{ojebode-2009-portrayal-europe}. Pakistani and Indian newspapers copy AP, AFP and Reuters \cite{shaheen-2021-international-news,mitra-2016-social-construction}. Malaysia and India belong to the semi-periphery here, so single-country studies cannot show the difference of degree the account implies. Reliance also carries frames. \citet{paterson-1999-agency-source} traced framing set in agency planning into broadcasters' stories and expected smaller broadcasters to relay strictly what the agencies tell, and \citet{riegert-2011-pondering-future} expected television's dependence to persist as correspondents decline. In 92 countries, online journalism stood closest to the emergency imaginary, the master frame that presents humanitarian issues as emergencies, which dependence on AP, AFP and Reuters copy may explain \cite{wright-2025-beyond-emergencies}. This study calls that reliance structural dependence and measures it by wire credits and by the presence of an outlet's own correspondents abroad.

\subsection{Choice built by structure}

What looks like choice may itself be a product of structure, because newsroom norms were exported. Media imperialism, the control over the means and contents of communication that comes with imperial power, remains a useful concept \cite{boydbarrett-2023-media-imperialism_abs}, and \citet{lee-1988-communication-imperialism} counted the transfer of broadcasting norms and professional models among its aspects. Similarities in television news worldwide have been attributed to Western practice serving as the professional model \cite{hallin-2004-americanization-globalization}, dependence on the agencies cultivates shared norms and formats \cite{riegert-2011-pondering-future}, and South African online media that rely on Western sources export Western norms \cite{radebe-2022-features-media_abs}. One channel is documented. Many Afghan journalists were trained by Western organizations funded by the United States, and Afghan and Pakistani journalists ``have a complex attachment to the American news media, which they scour for facts and insight, but also resent'' \cite[p.~12]{brown-2013-patterns-chaos}. Where a state agency carries the news, dependence may run to it instead. The state drives information circulation in the global peripheries \cite{abreu-2018-agencias-noticias}, national agencies are a staple of Global South media that research has neglected \cite{paterson-2026-researching-news_abs}, and the semi-periphery deserves study in its own right \cite{barry-2012-notes-semi}. This study does not measure how norms travel. It records credits to national agencies apart from the three Western wires and samples both tiers.

\subsection{Three rival accounts}

The rival inside the structural camp is national interest. \citet{skurnik-1981-look-foreign} stated the premise of the debate over the imbalance of world news, that newspapers in developing countries which depend on Western wires have little choice but to print what they get, and then tested it with Kenyan evidence. He judged the dependence explanation an oversimplification, because foreign news is selected under national interests. Audiences, too, seek cultural proximity, content close to their own culture \cite{straubhaar-1991-beyond-media}. Both predict less following on issues of high local stake, which the structural account denies where resources are thinnest.

The first voluntary rival is prestige imitation. Institutional theory holds that organizations in one field grow alike \cite{dimaggio-1983-iron-cage_abs}. DiMaggio and Powell called copying under uncertainty mimetic isomorphism and copying through shared professional norms normative isomorphism. Public service media in six European countries relied on what they considered best journalistic practice to reduce uncertainty \cite{sehl-2021-public-service}, news leaders may not accept stories until prestigious organizations have reported them \cite{brown-2013-patterns-chaos}, and Western European media systems have grown alike in products and practices \cite{hallin-2004-americanization-globalization}. Unlike herding, prestige imitation copies particular peers for their standing, not whoever moved first. It predicts that peripheral outlets copy the core's frames even without wire copy. Its evidence comes from within national media systems.

The second voluntary rival is herding. \citet{bikhchandani-1992-theory-fads_abs} showed that it can be optimal to follow those ahead without regard to one's own information, an informational cascade, and \citet{wilczek-2016-herd-behaviour} applied such herd behavior to journalists allocating attention. Directly competing outlets go into intense coverage of the same story together \cite{boydstun-2011-testing-explaining}, and in an agent-based model, surges of coverage gain momentum mainly from journalists adapting to what others cover \cite{waldherr-2014-emergence-news}. Herding predicts that following is fastest when a story is new and uncertain, because early movers are then all an outlet can observe. None of these studies tests herding across borders. Platform attention is a boundary of this study, not a rival. A news portal fed by Western agencies carried agency influence rather than replacing it \cite{watanabe-2013-western-perspective_abs}.

\subsection{Three levels of following}

Salience is the importance the media give an issue, signaled by how much coverage it gets \cite{mccombs-1972-agenda-setting_abs}. Framing is the angle taken. Timing is the order in which outlets take a story up. Across borders, \emph{The New York Times} and \emph{The Guardian} showed intermedia agenda setting with Egyptian news but no dominant agenda \cite{mohammed-2021-intermedia-agenda}. Timing has been traced mainly within countries \cite{golan-2006-inter-media}, language barriers suppressed influence among Belgian media \cite{vliegenthart-2008-contingency-intermedia_abs}, and most intermedia studies stay at salience \cite{su-2021-mapping-intermedia_abs}. \citet{harder-2017-intermediaagendase} traced copying to cost, since replicating others' content needs fewer resources, and to co-orientation, the habit of letting peers' coverage guide one's own. The structural account predicts following at all three levels that rises with wire copy. Prestige imitation predicts frame-following that remains without wire copy. Herding predicts timing following strongest for new, uncertain stories.

\subsection{Why a simulation}

A simulation can switch a mechanism off. Language-model agents are seen as a promising avenue for agent-based modeling, with open challenges in perception, alignment and evaluation \cite{gao-2024-large-language_abs}. \citet{park-2023-generative-agents_abs} built generative agents, which store their experiences in natural language and retrieve them to plan behavior. \citet{trnberg-2023-simulating-social} compared news-feed algorithms on simulated platforms. \citet{johansen-2025-imitating-agents} simulated imitation on all Danish digital news of 2022 and found that its effects depend on the information environment, which argues for calibrating a simulation to the outlets and issues modeled. The method has known limits. \citet{larooij-2025-large-language_abs} found validation of such models poorly addressed and argued that language models, whose inner workings cannot be inspected, are of limited use for disentangling causal mechanisms. \citet{yan-2026-fidelity-large} found language-model agents faithful to the direction of human agenda-setting effects but not to their magnitude. This study answers in part. The mechanisms are explicit rules on the model's proposals, judged against held-out coverage rather than believability, and a simulation can show only that a mechanism is sufficient to produce the pattern. The gap follows. Studies of the periphery measure reliance on agencies. Studies of national media systems measure imitation among peers. News-flow studies set national interest against dependence with coverage counts. Simulations vary imitation within one country. None sets structural dependence against prestige imitation and herding for the same outlets, issues and period, and the accounts share a prediction, so a correlation cannot tell them apart.

\subsection{Research questions and hypotheses}

The first question is descriptive, since intermedia studies have found a leading role for high-power countries beside a case with no dominant agenda. It establishes how much following there is before the hypotheses ask what explains it. RQ1. How closely do periphery and semi-periphery outlets follow core outlets on the salience, framing and timing of three global issues? The structural account's premise is that newsrooms lower in the world-system have fewer resources to gather foreign news. Heavy wire use is documented in Malaysia and India as well, so the premise predicts a difference of degree. H1. Periphery outlets follow core outlets more closely than semi-periphery outlets do. The second prediction concerns the channel. Agency planning shaped broadcasters' framing and newspapers without correspondents copy agency content, but no study has measured a gradient across outlets. H2. The more an outlet relies on wire copy for foreign news, the more closely it follows core outlets. The third sets the account against national interest and cultural proximity, which predict less following where local stake is high, while the account predicts that following persists where resources are thinnest. Skurnik's Kenyan evidence cuts against that premise. H3. Local stake in an issue reduces following less among periphery outlets than among semi-periphery outlets. The last question sets the mechanisms against each other, since each predicts the correlations observational studies report. National interest is not simulated, because the issues carry it and H3 tests it. Normative isomorphism is not simulated, because a norm every outlet shares cannot be switched off for some. RQ2. In the simulation, does structural dependence reproduce the observed following better than prestige imitation or herding? A win for dependence gives the resource account a test it has lacked. A win for either rival would show imitation or cascade sufficient across borders and call for revising it.

\section{Method}

\subsection{Type of study}

The study combined computational content analysis with an agent-based simulation. Computational content analysis uses software rather than people to read and classify large collections of news texts, which cuts the cost of analysis but needs problem-specific validation \cite{grimmer-2013-text-as-data}. An agent-based simulation is a computer model in which each outlet is a separate decision-maker that follows explicit rules. Neither alone can separate the mechanisms. A correlation between agendas does not show which mechanism produced it, and a simulation can turn a mechanism off. Language models served as the coder in the first part, as in recent annotation studies \cite{pangakis-2023-automatedannotatio,ziems-2024}, and as the agents in the second. Scholars disagree on whether such agents can reveal a mechanism \cite{gao-2024-large-language_abs,larooij-2025-large-language_abs}. The study meets the objection in three ways. The mechanisms were explicit rules acting on the agents' proposals, the rules were fitted on one period and scored on another, and the coder was validated by a stability protocol described under Validation. The data were published news texts, their metadata and model outputs.

\subsection{Data and sampling}

The study sampled six countries, Bangladesh, India, Kenya, Malaysia, Nigeria and South Africa, and 18 English-language national outlets, three per country. The countries were chosen because each has a national English-language press and because together they place South Asia, Africa and Southeast Asia in both tiers, so that the tier contrast is not a contrast between regions. In each country the three general-interest national outlets with the most stories over the period in the national collection of Media Cloud were taken, provided the archive held their stories in at least 80 percent of weeks. Media Cloud is an open archive of online news with a collection for each of the six countries \cite{mediacloud-2021}. Story records, meaning the date, title and URL of every story from February 24, 2022, to December 31, 2024, and each outlet's weekly count of all stories, came from Media Cloud. The GDELT Global Knowledge Graph, an open index of news events and story URLs worldwide that has served to map the news produced in 67 countries \cite{gdelt-2013,guovargo2017-global-intermedia}, served as the cross-check and as the source of event records. The core set for following was \emph{The New York Times}, \emph{The Washington Post}, \emph{The Guardian} and BBC News. AP, AFP and Reuters were sampled as the source of wire copy, not as outlets to follow.

A country's position was scored with the index of \citet{gunaratne-2001-prospects-limitations}, which weights a country's share of world gross national product at 55 percent and its share of world exports at 45 percent, from World Bank data for 2022 \cite{worldbank-2024-wdi}. Nigeria's 2022 exports came from the World Bank balance-of-payments series, because the national-accounts series was empty for that year, and Nigeria outranked South Africa under every index variant the authors checked. Because Gunaratne found the cutting point of the semi-periphery hard to fix, the three higher-scoring countries formed the semi-periphery, a declared departure from taking tiers from a published classification. The scores are reported, and position also entered the H1 models as a continuous score. A story belonged to one of three issues, the war in Ukraine, the war in Gaza or the COP27 and COP28 climate summits, when its title or text matched the issue's query terms (Appendix Table~\ref{tab:S1}). Duplicates, the same URL or at least 90 percent text overlap within an outlet, were removed. For framing and wire credits a random sample of up to 50 news articles per outlet and issue was drawn, core outlets included. Opinion pieces were excluded by section label, and an article whose public text could not be retrieved was replaced by the next draw. Frames were coded from the headline and lead paragraph, the window available for every outlet, including those behind paywalls. An outlet-issue cell with fewer than 20 stories was dropped. Nothing behind a login was collected, except the \emph{New York Times} and \emph{Washington Post} leads read in library databases under their terms of use.

Media Cloud's Bangladesh collection did not hold the leading English dailies, so \emph{The Daily Star}, \emph{bdnews24.com} and \emph{New Age} were chosen purposively and their series came from GDELT page titles. The same title-based measure was computed for every outlet as a check. \emph{The Washington Post} dropped out of the core series after March 4, 2024, and its 49 syndicated copies stayed as headline-only records credited as wire copy by URL. The \emph{Times of India}'s weekly count was Media Cloud's domain total scaled by the newspaper's share of that domain in GDELT. Sampled texts came from the outlets' public pages, from Internet Archive copies for two outlets and from library databases for the two paywalled core leads.

\subsection{Measures}

Following was computed for each outlet and issue at three levels, salience, framing and timing, for the whole period and for each validation half. Salience following is the correlation between an outlet's weekly share of stories on an issue and the core set's weekly share, after the weekly count of GDELT events for the issue was regressed out of both series, so that an outlet and the core both reacting to the same events did not count as following. A value of 1 means the outlet's attention rose and fell exactly with the core's, and 0 means no relation. It used the weeks in which both series held at least 5 stories and needed at least 8 such weeks. Framing following measures how far an outlet's mix of story angles sat from the core's. Frames were generic categories, such as the episodic and the conflict frame \cite{camaj-2010-media-framing}. The gap between the two frame distributions was the Jensen–Shannon distance, a standard measure of how different two probability distributions are \cite{lin-1991-divergence}. Following was 1 minus that distance, minus the value expected when both samples are redrawn from the pooled articles at the same sizes, a correction for the apparent difference small samples produce. So 0 means the outlet's frames are indistinguishable from the core's and a negative value means farther apart than sampling alone would give. Timing following is the median lag in days between the first core story and the first outlet story on the same event, with events matched to stories through the article URLs GDELT links to its event records. It is entered as a negative number, so a less negative value means closer following, and an event not covered within 14 days counted as 14, a censored value, in either direction. A cell needed at least 10 matched events. Framing was also recomputed without the core stories that credit a wire, a check possible for framing only, because the core's census records carry no credit codes.

Wire reliance is the share of an outlet's sampled foreign stories, those datelined abroad or credited to a foreign source, that credit AP, AFP or Reuters in the byline, dateline or end note. Credits followed fixed rules released with the codebook. No text-reuse measure was applied, because the agencies' full texts were not collected, so uncredited copy is missed. Credits to national agencies were recorded apart. Correspondent presence is the share of foreign stories filed by an outlet's own staff under a foreign dateline. It fell below the reliability floor, so it entered no confirmatory test, but it still entered the simulation's dependence rule, which was fixed before the data. Local stake was fixed by issue, Ukraine 0, Gaza 1 and the climate summits 2. No country-by-issue stake was coded, since nothing in the study was hand-coded. The share of an outlet's issue stories that name its own country served as an outlet-level check.

\subsection{Simulation design}

The simulation had to let each mechanism be switched on or off while the rest stayed fixed, and to reproduce observed following without seeing it. It adapted Magentic Marketplace, an open environment in which some agents represent buyers and others sellers \cite{gagan-2025-magentic-marketplace_abs}. The three wire agencies supplied copy, the four core outlets replayed their observed stories, and the 18 peripheral outlets were agents run on Qwen3-14B, a language model whose weights are public, served with the vLLM software on Google Colab GPUs. Each day, each agent read a feed of GDELT event records, wire stories and core stories at their observed times, stripped of outlet names, and recorded its proposal, the items it would publish, each with a priority and a frame from the codebook. No observed story of a peripheral outlet entered the feed. Each agent's persona gave only a country and an audience, and each agent read its own proposals of the previous seven days as memory \cite{park-2023-generative-agents_abs}. Proposals were recorded from two feeds, one with core stories and one without, three samples per outlet-day. Because the model proposed once, agents did not adapt to what a rule had done on earlier days. In return the rules were exact, which narrows the concern that language models obscure causal mechanisms \cite{larooij-2025-large-language_abs} to the proposals.

Each mechanism was a rule with one free parameter acting on the proposals (Appendix Table~\ref{tab:rules}). Structural dependence capped the number of an outlet's own foreign stories on a day. The outlet kept its highest-priority non-wire proposals up to the cap and took the rest of its foreign news from the day's wire copy and frames, following the premise that wire services form a finite basis for editorial choice \cite{skurnik-1981-look-foreign}. Prestige imitation pulled the weight of each proposed issue and frame toward its share among the core outlets that day, modeling organizations that copy peers they see as more legitimate \cite{sehl-2021-public-service}. Herding let an outlet publish a new, uncertain story, whether or not it had proposed it, with a probability that rose with the number of outlets that had already covered it, and adopt the frame most earlier adopters used, following an informational cascade \cite{bikhchandani-1992-theory-fads_abs}. A story was new within 2 days of its first core story and uncertain while fewer than 3 core stories covered it. Each mechanism was run alone, each pair and all three together, plus a baseline that published the proposals with no rule, eight configurations, each 20 times with different random draws, or seeds. Combined rules applied in the order dependence, imitation, herding. Only the dependence rule read outlet-specific inputs, because only the structural account predicts that outlets differ by their resources.

\subsection{Validation}

Two kinds of validation are reported. First, no variable was hand-coded and no human verified a sample. Credits, datelines and bylines followed fixed rules. The other codes came from the model coder, a Claude model whose version was recorded before coding. Its reliability was measured by a stability protocol. The coder labeled the same random sample of 300 articles three times in fresh contexts, and agreement among the passes was summarized as Krippendorff's alpha, a statistic that runs from 0 (agreement no better than chance) to 1 (perfect agreement), on the finding that labels on which repeated runs agree are more accurate than labels on which they differ \cite{pangakis-2023-automatedannotatio}. A floor of .80, set before coding, applied to that across-pass alpha and to agreement between the fixed rules and the coder. The alpha against a reworded prompt was reported beside them as a sensitivity figure. No human spot-check was run, so stability shows consistency, not validity \cite{ziems-2024}. A variable below the floor after one revision became exploratory. The agents labeled their proposals' frames with the same codebook. For 300 proposals they also wrote a headline and lead that the coder coded, and agent labels had to reach the same floor, otherwise simulated framing used the coder's labels on the written-out proposals. Before the proposal pass, agents were asked to recall headlines of the sampled outlets for sampled dates, a contamination check for whether the model had memorized the coverage. Second, rule parameters were set on the first half of each issue and scored on the second, COP27 against COP28 for the climate issue and date halves for each war, so that fit was judged on data the rules had not seen \cite{larooij-2025-large-language_abs}.

\subsection{Analytic strategy}

Each hypothesis had a procedure and a decision rule set before the data. RQ1 was answered descriptively by tier and issue, against a benchmark computed from the core series of a different issue, which shows what following looks like when there is none. H1 to H3 were tested with mixed models, regressions that allow outlets within the same country to share an unmeasured tendency, fitted separately for salience, framing and timing with random intercepts for outlet and country and issue as a fixed effect. A hypothesis was supported at a level if its coefficient had the predicted sign and a p value below .05 after Holm's correction, an adjustment that raises each p value to allow for the nine tests run at once. A 95\% confidence interval is reported beside each estimate, and because three countries per tier give little power, a null for H1 is read from its interval, not as evidence of no difference. H2 was tested in a model that kept tier, which also read the account's first falsifier, a result that would count against it. If the tier gap net of wire reliance had an upper bound below half the H1 estimate, the account's other channels, correspondents and capital, would lose support. The falsifier was read only where the H1 estimate was positive. H3 replaced issue with stake and was supported if following dropped less from low to high stake in the periphery. Its falsifier held if periphery following at high stake was no greater than the mismatched-issue benchmark. Two checks repeated H1 with the continuous index score and H3 with the own-country share. The rivals also got observational tests, decided by Holm across the pair. For prestige imitation, framing following was modeled on stories without wire copy, where a positive intercept counted as support. Because the bias-corrected similarity has an expected value of at most zero, this rule can fail to support but cannot support. For herding, within each war, the negative first-story lag on an event was regressed on the event's uncertainty, measured as the number of core outlets that mentioned it on its first core day, where a negative coefficient supports herding. Every analysis not listed here is exploratory.

RQ2 was answered by out-of-sample error. For each level, a configuration's error was the mean absolute difference between simulated and observed following across cells, divided by the standard deviation of the observed values, so that 1.00 means the simulation missed by one observed spread on average. Its overall error was the mean over the three levels. A mechanism won if its single-mechanism configuration had the lowest overall error and beat the other two and the baseline by more than twice the across-seed standard deviation of the seed-matched difference, the seed margin, and if that margin held in at least 95 percent of bootstrap resamples of outlets, re-draws of the 18 outlets with replacement, so that a win reflected the observed targets and not a few outlets. As a check, error was also computed within outlets, on each outlet's deviations from its own mean across issues and halves, where outlet-specific inputs give dependence no advantage. The sign of each observed tier and stake contrast was reported beside each error, since language-model agents reproduce direction better than magnitude \cite{yan-2026-fidelity-large}. RQ2 is undecided if no mechanism wins by the margin.

\subsection{Reproducibility}

A public repository at \url{https://github.com/hellowyousuf/dependence-imitation-herding} holds the adapted environment, the rule code, the codebook, the prompts and the proposals. Its release, archived on Zenodo under DOI \href{https://doi.org/10.5281/zenodo.23330238}{10.5281/zenodo.23330238}, holds every script, an environment file, fixed seeds and a README listing the command, runtime and hardware for each result. Because the rules act on the released proposals, the simulation analysis can be rerun without a language model. Article texts are not released, since they belong to the outlets. Generative AI assisted literature screening and drafting under the authors' review. The tools and their roles are listed in the Declaration of generative AI use.

\section{Results}

Both tiers followed the core's attention, matched its timing within a day and sat closer to its frames than the benchmark but not on them (RQ1). H1 was not supported, H2 was supported for timing only, and H3 was not supported. The isomorphism test was not supported and the herding test was supported. RQ2 is undecided. The numbers follow, validation first.

\subsection{Validation}

Every confirmatory variable met the stability floor, and no human coded any story (Appendix Table~\ref{tab:S3}). The frame code reached a Krippendorff's $\alpha$ of .849 across the three passes and .799 against a reworded prompt, the sensitivity figure outside the floor. The foreign-story code reached .956 after one revision of its wording but only .406 against the reworded prompt, and the fixed rule and the coder agreed at 1.000 on the 1,091 stories the rule decided. The wire-credit rule was stable at .995 and agreed with the coder at .996 over 3,072 stories, the national-agency rule at 1.000 and .991. Staff foreign bylines fell below the floor at .749 and stayed exploratory. The agents' own frame labels agreed with the coder at only .232, so simulated framing used the coder's labels on written-out proposals. In the contamination check the agents recalled none of 100 sampled headlines at a similarity of 0.8 or more and were no closer to the true headline (0.270) than to a random one (0.273).

\subsection{Following by tier (RQ1)}

\begin{table*}[tbp]
\centering
\footnotesize
\caption{Following at Three Levels by Tier and Issue}
\label{tab:following}
\setlength{\tabcolsep}{3pt}
\begin{tabularx}{\textwidth}{>{\raggedright\arraybackslash}p{0.09\textwidth}>{\raggedright\arraybackslash}p{0.08\textwidth}>{\raggedright\arraybackslash\hsize=0.873\hsize}X>{\raggedright\arraybackslash\hsize=0.873\hsize}X>{\raggedright\arraybackslash\hsize=1.091\hsize}X>{\raggedright\arraybackslash\hsize=1.091\hsize}X>{\raggedright\arraybackslash\hsize=1.200\hsize}X>{\raggedright\arraybackslash\hsize=0.873\hsize}X}
\toprule
\textbf{Tier} & \textbf{Issue} & \textbf{Cells (n, after the 20-story floor)} & \textbf{Coded articles (n)} & \textbf{Salience r (M, SD)} & \textbf{Framing similarity (M, SD)} & \textbf{Timing: first-story lag (Mdn days)} & \textbf{Timing: peak cross-correlation lag (Mdn weeks)} \\
\midrule
Periphery & Ukraine & 9 & 433 & 0.33 (0.19) & $-$0.12 (0.09) & $-$0.12 (events 9,167) & 0 \\
Periphery & Gaza & 9 & 449 & 0.19 (0.23) & $-$0.10 (0.11) & $-$0.04 (events 10,814) & 0 \\
Periphery & COP27/COP28 & 8 & 364 & 0.64 (0.29) & $-$0.09 (0.09) & 0.04 (events 131) & 0 \\
Periphery & All issues & 26 & 1246 & 0.38 (0.29) & $-$0.10 (0.09) & $-$0.07 (events 20,112) & 0 \\
Semi-periphery & Ukraine & 9 & 448 & 0.28 (0.27) & $-$0.08 (0.06) & $-$0.04 (events 11,530) & 0 \\
Semi-periphery & Gaza & 9 & 450 & 0.11 (0.30) & $-$0.11 (0.07) & $-$0.03 (events 14,288) & 0 \\
Semi-periphery & COP27/COP28 & 8 & 385 & 0.46 (0.30) & $-$0.09 (0.09) & 0.00 (events 159) & 0 \\
Semi-periphery & All issues & 26 & 1283 & 0.27 (0.31) & $-$0.09 (0.07) & $-$0.03 (events 25,977) & 0 \\
\bottomrule
\end{tabularx}
\par\smallskip
\parbox{\textwidth}{\raggedright \emph{Note.} Salience r is the Pearson correlation between the event-adjusted residuals of an outlet's weekly issue share and the core set's over the weeks both hold at least 5 stories (149 weeks, February 24, 2022, to December 31, 2024). The core set is the weekly mean of \emph{The New York Times}, \emph{The Washington Post}, \emph{The Guardian} and BBC News over the core outlets present that week; \emph{The Washington Post} is absent after March 4, 2024. Framing similarity is 1 minus the Jensen–Shannon distance between the cell's frame distribution and the pooled core sample's, minus its expected value over 1,000 random splits of the pooled articles at the observed sizes, so that 0 means indistinguishable from the core and negative values mean farther than sampling alone would give. The first-story lag is the negative median lag in days from the first core mention of a GDELT event to the outlet's first mention; lags above 14 days are censored; the peak cross-correlation lag is the negative lag in weeks of the peak cross-correlation of weekly shares ($-$4 to 4). Dropped cells (fewer than 20 stories): \emph{Malaysiakini} COP27/COP28, \emph{New Age} COP27/COP28. Cells with 20–49 stories were coded whole: \emph{bdnews24.com} COP (33), \emph{New Sarawak Tribune} COP (36), \emph{Tuko} COP (33). The first-story lag is reported only for cells with at least 10 matched events (37 of 52 kept cells); \emph{New Sarawak Tribune} has no GDELT event mentions, so its three cells have no first-story lag. Tiers by Gunaratne's index on World Bank 2022 data: periphery = Bangladesh, Kenya, South Africa; semi-periphery = India, Malaysia, Nigeria.}
\end{table*}

\begin{figure}[tbp]
\centering
\includegraphics[width=\columnwidth]{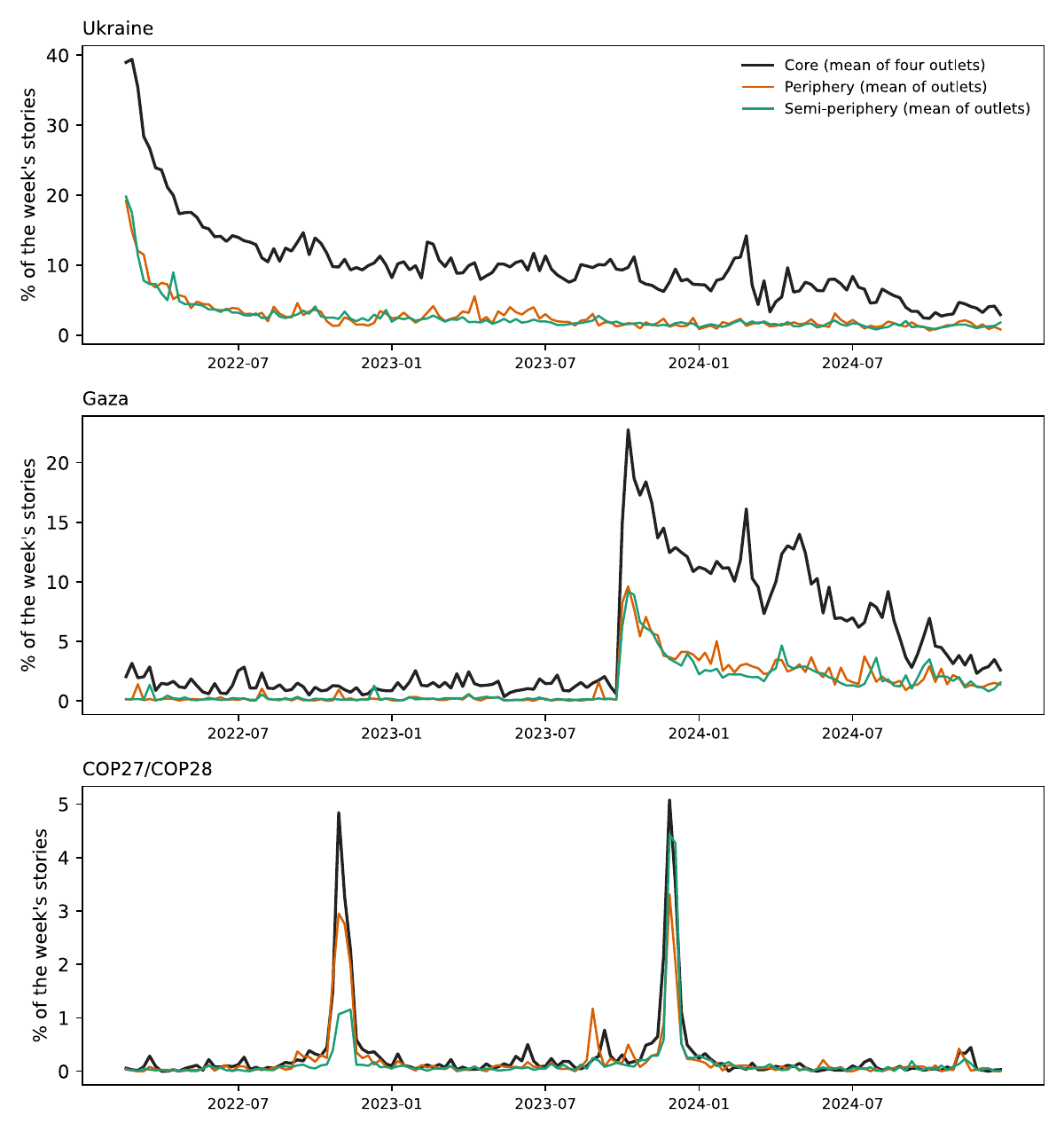}
\caption{Weekly Issue Shares, Core and Peripheral Outlets}
\label{fig:shares}
\par\footnotesize\parbox{\columnwidth}{\raggedright \emph{Note.} Weekly share of each outlet’s stories on the issue, core set (mean of the four core outlets present that week) against the peripheral outlets by tier, February 24, 2022, to December 31, 2024. Source: run\_20261008\_02.}
\end{figure}

The answer to RQ1 is that both tiers followed the core's attention and timing and sat just outside its frames (Table~\ref{tab:following}; Figure~\ref{fig:shares}). Two cells fell below the 20-story floor, leaving 52. Over the 26 periphery cells the event-adjusted salience correlation averaged 0.38 (SD = 0.29) and over the 26 semi-periphery cells 0.27 (SD = 0.31), against a benchmark of $-$0.13 and $-$0.15 from the core series of a different issue. Attention following was highest on the climate conferences (0.64 and 0.46) and lowest on Gaza (0.19 and 0.11). Framing similarity averaged $-$0.10 (SD = 0.09) and $-$0.09 (SD = 0.07), above the benchmark of $-$0.16 and $-$0.14 but below the zero that would mean indistinguishable frames. The median first-story lag was $-$0.07 days in the periphery and $-$0.03 in the semi-periphery, within two hours of the core's first mention. Its benchmark, the core's first mentions shuffled across events, was 0.30 (SD = 4.68) and $-$0.07 (SD = 2.16). Timing existed for 37 of the 52 cells. Credited wire copy carried most of the periphery's foreign news. AP, AFP or Reuters credits covered 0.57 of the periphery's 1,017 coded foreign stories and 0.40 of the semi-periphery's 823, ranging from 0.00 at \emph{New Age} to 0.94 at \emph{Malay Mail}, against 0.14 in the core. Staff foreign bylines were 0.00 and 0.01, and national-agency credits reached 0.12 in the semi-periphery, concentrated in two Malaysian outlets (Appendix Table~\ref{tab:S4}). Two checks qualify the salience figures (Appendix Tables~\ref{tab:S5} and~\ref{tab:S6}). Salience following was about half as strong in the second validation half (0.22 and 0.23) as in the first (0.41 and 0.44). Recomputed from GDELT title-based shares it correlated with the main values at only 0.70 in the periphery and 0.18 in the semi-periphery, with tier means of 0.28 and 0.30 that remove the descriptive gap, so the salience measure depends on the source more than the weekly shares do (r = 0.68 and 0.87).

\subsection{Hypothesis tests (H1–H3)}

\begin{table*}[tbp]
\centering
\footnotesize
\caption{Mixed Models of Following on Tier (H1), Wire Reliance (H2) and Tier $\times$ Local Stake (H3), With the Holm Decisions and the Two Falsifiers}
\label{tab:models}
\setlength{\tabcolsep}{3pt}
\begin{tabularx}{\textwidth}{>{\raggedright\arraybackslash\hsize=0.833\hsize}X>{\raggedright\arraybackslash\hsize=1.204\hsize}X>{\raggedright\arraybackslash\hsize=0.556\hsize}Xr>{\raggedright\arraybackslash}p{0.11\textwidth}>{\raggedright\arraybackslash\hsize=0.926\hsize}Xr>{\raggedright\arraybackslash\hsize=1.481\hsize}X}
\toprule
\textbf{Test} & \textbf{Predictor} & \textbf{Level} & \textbf{b} & \textbf{95\% CI} & \textbf{p} & \textbf{Cells} & \textbf{Decision} \\
\midrule
H1 & Tier (periphery) & Salience & 0.103 & [$-$0.148, 0.354] & 1.000 & 52 & not supported \\
H1 & Tier (periphery) & Framing & $-$0.010 & [$-$0.065, 0.046] & 1.000 & 52 & not supported \\
H1 & Tier (periphery) & Timing & $-$0.026 & [$-$0.132, 0.081] & 1.000 & 37 & not supported \\
H2 & Wire reliance & Salience & $-$0.020 & [$-$0.267, 0.227] & 1.000 & 52 & not supported \\
H2 & Wire reliance & Framing & 0.019 & [$-$0.066, 0.104] & 1.000 & 52 & not supported \\
H2 & Wire reliance & Timing & 0.175 & [0.084, 0.267] & .002 & 37 & supported \\
H3 & Periphery $\times$ stake & Salience & 0.061 & [$-$0.114, 0.236] & 1.000 & 52 & not supported \\
H3 & Periphery $\times$ stake & Framing & 0.019 & [$-$0.032, 0.070] & 1.000 & 52 & not supported \\
H3 & Periphery $\times$ stake & Timing & 0.049 & [$-$0.020, 0.117] & 1.000 & 37 & not supported \\
H1 check & Index score (per point) & Salience & 0.025 & [$-$0.119, 0.168] & .738 (unadj.) & 52 & check \\
H1 check & Index score (per point) & Framing & 0.018 & [$-$0.011, 0.046] & .219 (unadj.) & 52 & check \\
H1 check & Index score (per point) & Timing & 0.007 & [$-$0.049, 0.064] & .794 (unadj.) & 37 & check \\
H3 check & Periphery $\times$ own-country share & Salience & 2.510 & [$-$0.916, 5.937] & .151 (unadj.) & 52 & check \\
H3 check & Periphery $\times$ own-country share & Framing & $-$0.068 & [$-$0.981, 0.846] & .885 (unadj.) & 52 & check \\
H3 check & Periphery $\times$ own-country share & Timing & 0.092 & [$-$1.731, 1.916] & .921 (unadj.) & 37 & check \\
First falsifier & Periphery, net of wire reliance & Salience & 0.108 & [$-$0.153, 0.368] & .418 (unadj.) & 52 & not met \\
First falsifier & Periphery, net of wire reliance & Framing & $-$0.014 & [$-$0.074, 0.047] & .654 (unadj.) & 52 & not applicable (H1 estimate negative) \\
First falsifier & Periphery, net of wire reliance & Timing & $-$0.053 & [$-$0.141, 0.034] & .232 (unadj.) & 37 & not applicable (H1 estimate negative) \\
Second falsifier & Mean following 0.64 vs benchmark $-$0.20 & Salience & 0.837 & paired t(7) = 9.57 & one-sided < .001, Holm < .001 & 8 & falsifier does not hold (following above benchmark) \\
Second falsifier & Mean following $-$0.09 vs benchmark $-$0.12 & Framing & 0.025 & paired t(7) = 1.31 & one-sided .116, Holm .116 & 8 & falsifier holds (following not above benchmark) \\
Second falsifier & Mean following 0.04 vs benchmark 0.04 & Timing & — & — & one-sided —, Holm — & 2 & not testable (too few cells) \\
\bottomrule
\end{tabularx}
\par\smallskip
\parbox{\textwidth}{\raggedright \emph{Note.} Cells are outlet $\times$ issue (52; timing 37 with at least 10 matched events). Mixed models with random intercepts for outlet nested in country and issue as a fixed effect (H3: stake in place of issue, coded Ukraine 0, Gaza 1, COP 2); a positive b means closer following in the periphery (H1), with more wire copy (H2) or a smaller stake-related drop in the periphery (H3). p for the nine confirmatory rows is Holm-adjusted across the nine tests; checks and falsifier read-outs show unadjusted p. Where the country variance was estimated at zero the model was refitted without it. Full model output, variance components and sensitivity fits without the headline-only Post copies are in Appendix Tables~\ref{tab:S9}--\ref{tab:S11}.}
\end{table*}

H1 was not supported at any level. Tier did not separate the two groups (Table~\ref{tab:models}). The periphery coefficient was 0.103 on salience (95\% CI [$-$0.148, 0.354]), $-$0.010 on framing (95\% CI [$-$0.065, 0.046]) and $-$0.026 on timing (95\% CI [$-$0.132, 0.081]), each with a Holm-adjusted p of 1.000. The check with the continuous index score, which predicts a negative coefficient, gave 0.025, 0.018 and 0.007 per index point, positive at every level with intervals that include zero. No interval is narrow enough to read the null as no difference. The framing interval is wider than the tier means' standard deviations of 0.09 and 0.07.

H2 was supported for timing only. Reliance on wire copy went with closer timing, not with closer salience or framing. On timing the wire-reliance coefficient was 0.175 (95\% CI [0.084, 0.267], 37 cells, Holm-adjusted p = .002). A cell with wire credits on every foreign story reached an event 0.175 days, about four hours, sooner than one with none. On salience it was $-$0.020 (95\% CI [$-$0.267, 0.227]) and on framing 0.019 (95\% CI [$-$0.066, 0.104]), Holm-adjusted p = 1.000 for both. The timing result describes the channel as much as a choice. An outlet that prints wire copy publishes at the wire's hour, so the lag closes whether or not anyone follows. The first falsifier applied on salience alone, where the tier estimate was positive. The tier gap net of wire reliance was 0.108 (95\% CI [$-$0.153, 0.368]) against a gross gap of 0.103, so the upper bound stayed above the 0.052 that would have cost the non-wire channels their support.

\begin{figure*}[tbp]
\centering
\includegraphics[width=\textwidth]{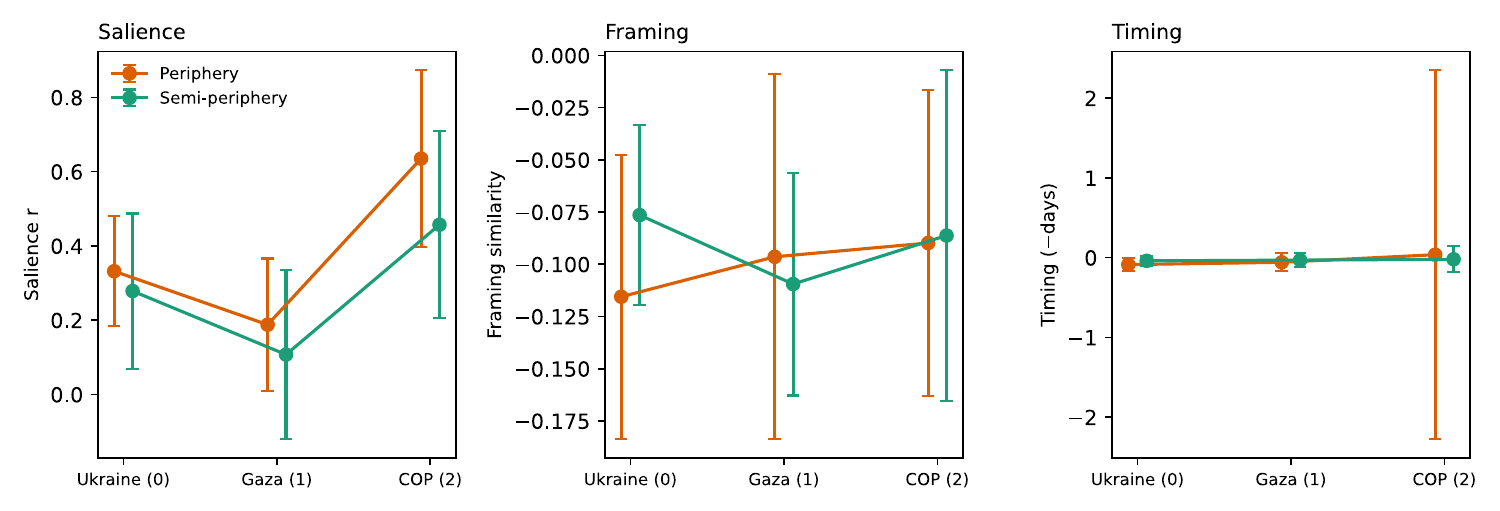}
\caption{Following by Tier and Local Stake}
\label{fig:stake}
\par\footnotesize\parbox{\textwidth}{\raggedright \emph{Note.} Cell means of salience, framing and timing following by tier at each stake level (Ukraine 0, Gaza 1, COP27/COP28 2), with the mismatched-issue benchmark. Source: run\_20261009\_01.}
\end{figure*}

H3 was not supported. Local stake did not reduce following less among periphery outlets (Figure~\ref{fig:stake}). Stake is coded at the issue level, one issue per level, so it is confounded with everything else that differs between the wars and the climate conferences. The periphery $\times$ stake interaction was 0.061 (95\% CI [$-$0.114, 0.236]) on salience, 0.019 (95\% CI [$-$0.032, 0.070]) on framing and 0.049 (95\% CI [$-$0.020, 0.117]) on timing, all positive as predicted, each with a Holm-adjusted p of 1.000. The own-country share in place of stake gave no interval excluding zero. The second falsifier compared the eight periphery climate cells with their mismatched-issue benchmark. On salience, following stood at 0.635 against $-$0.201 (paired t = 9.57, Holm-adjusted p < .001), so the falsifier did not hold and periphery attention followed the core where stake was highest. On framing it stood at $-$0.090 against $-$0.115 (paired t = 1.31, Holm-adjusted p = .116), so the falsifier held. On timing it was not testable.

\subsection{Rival tests}

The isomorphism test was not supported and the herding test was supported (Appendix Table~\ref{tab:rivals}). Framing similarity recomputed on stories without wire credits in both the outlets and the core had an intercept of $-$0.102 (95\% CI [$-$0.136, $-$0.068], 1,623 non-wire stories over 52 cells), which differed from the all-story similarity by $-$0.006. A positive intercept counted as support, so the test was not supported (Holm-adjusted p < .001 for the negative value), by a rule that could not have supported it. For herding, the negative first-story lag on an event was regressed on the number of core outlets that mentioned it on its first core day. The coefficient was $-$0.1316 days per additional core outlet in Ukraine (95\% CI [$-$0.1726, $-$0.0906], 20,697 events) and $-$0.1640 in Gaza (95\% CI [$-$0.1893, $-$0.1387], 22,665 events). Outlets followed sooner on the events the core had covered least, as a cascade predicts, and the test was supported (Holm-adjusted p < .001). Two readings this design did not test remain open. Events few core outlets mention may be the items wire copy delivers to every subscriber at once, and because the first core mention is the earliest of several, the lag to it grows with the number of core stories by construction.

\subsection{Simulation (RQ2)}

\begin{table*}[tbp]
\centering
\footnotesize
\caption{Out-of-Sample Fit of the Eight Simulation Configurations}
\label{tab:sim}
\setlength{\tabcolsep}{3pt}
\begin{tabularx}{\textwidth}{>{\raggedright\arraybackslash\hsize=1.346\hsize}X>{\raggedright\arraybackslash\hsize=0.962\hsize}X>{\raggedright\arraybackslash\hsize=0.962\hsize}X>{\raggedright\arraybackslash\hsize=0.962\hsize}X>{\raggedright\arraybackslash\hsize=0.962\hsize}X>{\raggedright\arraybackslash\hsize=0.962\hsize}X>{\raggedright\arraybackslash\hsize=0.962\hsize}X>{\raggedright\arraybackslash\hsize=0.962\hsize}X>{\raggedright\arraybackslash\hsize=0.962\hsize}X>{\raggedright\arraybackslash\hsize=0.962\hsize}X}
\toprule
\textbf{Configuration} & \textbf{Error: salience} & \textbf{Error: framing} & \textbf{Error: timing} & \textbf{Mean error} & \textbf{vs baseline (M, SD)} & \textbf{vs best other single rule (M, SD)} & \textbf{Signs reproduced} & \textbf{Bootstrap share} & \textbf{Within-outlet error} \\
\midrule
Baseline (no rule) & 1.13 & 1.06 & 0.79 & 1.00 & — & — & 0.50 & — & 0.93 \\
Dependence & 1.25 & 0.95 & 0.28 & 0.83 & $-$0.17 (0.01) & $-$0.08 (0.01) & 0.71 & 0.45 & 0.93 \\
Imitation & 0.85 & 1.08 & 0.79 & 0.91 & $-$0.09 (0.00) & 0.08 (0.01) & 0.50 & — & 0.88 \\
Herding & 1.22 & 1.09 & 1.69 & 1.33 & 0.34 (0.02) & 0.51 (0.02) & 0.53 & — & 1.39 \\
Dependence + imitation & 1.19 & 1.01 & 0.43 & 0.88 & $-$0.12 (0.01) & — & 0.44 & — & 0.88 \\
Dependence + herding & 1.30 & 0.98 & 7.29 & 3.19 & 2.20 (0.45) & — & 0.62 & — & 1.42 \\
Imitation + herding & 0.92 & 1.06 & 0.79 & 0.92 & $-$0.07 (0.01) & — & 0.46 & — & 0.89 \\
All three & 1.24 & 1.02 & 4.23 & 2.16 & 1.17 (0.23) & — & 0.50 & — & 0.89 \\
\bottomrule
\end{tabularx}
\par\smallskip
\parbox{\textwidth}{\raggedright \emph{Note.} Standardized error = mean absolute difference between simulated and observed following over outlet $\times$ issue cells on the second validation half, divided by the SD of the observed values; 20 seeds per configuration. Parameters fitted on the first half (w for dependence, v for imitation, c for herding): dependence w = 1.00; imitation v = 0.10; herding c = 0.50; each combination uses the same values for its rules, and a rule not in a configuration has its parameter at 0. Decision (seed margin and bootstrap condition, as the Method states): RQ2 undecided. Dependence met the seed margin, but its margin held in a share of 0.45 of outlet resamples, below the 0.95 the Method requires.}
\end{table*}

\begin{figure}[tbp]
\centering
\includegraphics[width=\columnwidth]{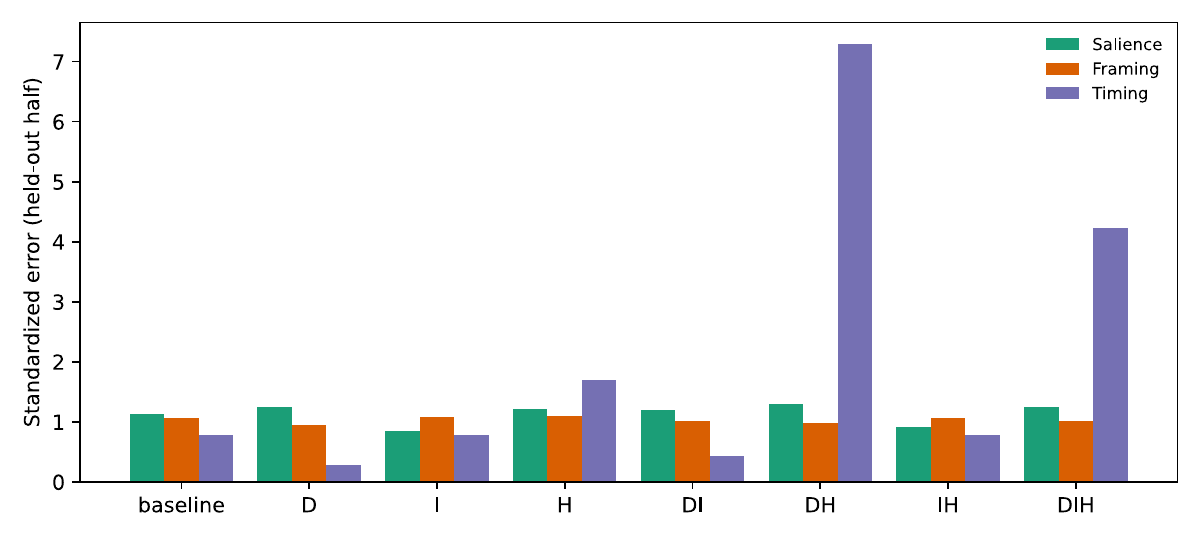}
\caption{Standardized Out-of-Sample Error by Level and Configuration}
\label{fig:simerror}
\par\footnotesize\parbox{\columnwidth}{\raggedright \emph{Note.} Mean absolute error between simulated and observed following on the second validation half, divided by the SD of the observed values, for each configuration and level; 20 seeds. D = dependence, I = imitation, H = herding; joined letters are combinations (DIH = all three). Source: run\_20261012\_01.}
\end{figure}

The answer to RQ2 is that no mechanism won under the Method's rule, so the comparison is undecided (Table~\ref{tab:sim}; Figure~\ref{fig:simerror}). Grid search on the first validation half, trying each value of a rule's dial and keeping the best, gave w = 1.00 for dependence, v = 0.10 for imitation and c = 0.50 for herding, two of them at the ends of their grids. On the second half the mean standardized error was 1.00 for the baseline, 0.83 for dependence, 0.91 for imitation, 1.33 for herding, 0.88 for dependence with imitation, 3.19 for dependence with herding, 0.92 for imitation with herding and 2.16 for all three. Dependence beat the baseline by 0.17 and the best other single mechanism by 0.08, each with a seed-matched SD of 0.01, so the seed margin was met. But the margin held in only 0.45 of the outlet resamples, below the 0.95 the Method requires, so RQ2 is undecided. Two read-outs show why (Appendix Tables~\ref{tab:S7} and~\ref{tab:S8}). At w = 1.00 the dependence cap reduced to n $\times$ (1 $-$ W) $\times$ C, and because correspondent presence C was 0.00 for 15 of the 18 outlets and at most 0.05 for the rest, 17 of 18 outlets kept no own story at all and published wire copy only. Wire copy carries no first-mention lag, so dependence alone held a timing value in 2.0 of the 52 cells per seed, against 52 for the configurations without it, and the timing error of 0.28 that carried its margin rests on two cells. Within outlets, dependence's error (0.93) equaled the baseline's. Imitation was the only single rule that lowered the salience error (0.85 against 1.13).

\section{Discussion}

\subsection{Structural dependence}

Structural dependence showed in the channel at one level, and the simulation did not decide between the mechanisms. The account predicted following strongest in the periphery, rising with wire reliance and persisting where local stake is high. The record supports the second prediction at the timing level. A cell with AP, AFP or Reuters credits on every foreign story reached events about four hours sooner than one with none. \citet{paterson-1999-agency-source} expected smaller broadcasters to relay strictly the agencies' stories. This study finds that relay in the clock, and at the attention and framing levels wire reliance showed no supported effect. Even the timing result has a reading this design cannot exclude. Wire copy reaches outlet and core at the same hour, and the lag closes without anyone following. The design shows association, not how newsrooms decide. The persistence reading of the third prediction held at the attention level, where periphery outlets followed the core on the climate conferences, the highest-stake issue, far above their benchmark, and not at the framing level, so the domestication of frames stays open.

The first prediction was a test the measure could not carry. Tier is a distant proxy for the resource the account names. The index placed Nigeria in the semi-periphery and South Africa in the periphery on a difference of 0.03 index points, while wire reliance ran from none of the foreign stories at one Bangladeshi daily to nearly all at one Malaysian outlet, with means over outlet-issue cells of 0.56 and 0.34 by tier and standard deviations of 0.37 and 0.32. Timing following tracked the share of wire copy while the country's score showed nothing. Staff foreign bylines were almost absent in both tiers, so, if that exploratory measure is near the truth, the outlets differed in how much agency copy they used, not in whether they had an alternative \cite{camaj-2010-media-framing}. The first falsifier was not met on salience because the interval stayed wide, not because a non-wire channel was shown.

\subsection{The rival accounts}

The test of prestige imitation where it predicts most was uninformative. Institutional theory expects frames to follow the core's where no wire copy carries them \cite{dimaggio-1983-iron-cage_abs}. The frames of stories without wire credits were no closer to the core's than the frames of all stories, both below what sampling alone would give, and the decision rule could not support, so the test decided nothing. The informative comparison is the benchmark. The periphery's frames were closer to the core's on the same issue than on a different one, a pattern in which the issue, more than the source, fixes the frame, as the frames \citet{camaj-2010-media-framing} found shared across agencies on Kosovo's status negotiations and the emergency master frame \citet{wright-2025-beyond-emergencies} found in agency copy would produce. Normative isomorphism, the shared professional norm that would make every outlet frame alike, was not simulated. The uniform framing distance across tiers and credits fits it without testing it. Herding kept its observational result, consistent with a cascade \cite{bikhchandani-1992-theory-fads_abs,wilczek-2016-herd-behaviour}, but the two untested readings stated in the Results, wire delivery and the order-statistic artefact, keep it at consistency. The simulation does not settle it, since herding as coded did worse than no rule, which says nothing about the gradient. The record reaches the cost reason for copying, not the co-orientation \citet{harder-2017-intermediaagendase} set beside it, and the competition account of \citet{boydstun-2011-testing-explaining} remains open. National interest \cite{skurnik-1981-look-foreign} and cultural proximity \cite{straubhaar-1991-beyond-media} did not reduce periphery attention to the climate conferences. Where an outlet credits a national agency, as two Malaysian outlets did on more than half of their foreign stories, a second pole of dependence exists that this design did not model.

\subsection{Limitations}

No human coded any story, and no human spot-check was run. The stability protocol shows that the coder is consistent with itself and with the fixed rules, and the foreign-story code reached only .406 against a reworded prompt, so stability is not validity. The data carry several compromises. The tiers rest on a 0.03-point difference and a substitute export series. Bangladesh's outlets were chosen purposively and their series drawn from GDELT titles, and the source check shows that salience following is sensitive to that choice. Malaysia's entered by a rule that admitted a Sarawak-based paper and a thin outlet. The \emph{Post}'s syndicated copies stayed as headline-only records. Wire reliance counts credits only, so uncredited copy is missed in an amount the data cannot show. Four outlets, \emph{New Age}, \emph{The Star} of Kenya, \emph{Daily Post} and \emph{Nigerian Tribune}, credited no wire at all, which may mean no wire copy or copy without credit, and correspondent presence, the dependence rule's input, is read from foreign datelines in the sampled stories, not from a check of each outlet's staff. Stake is an issue-level code with one issue per level. Timing exists for 37 of 52 cells, and all outlets are English-language, which leaves the vernacular press out. With 52 cells, the intervals around the tier and stake coefficients are wide, so the nulls are absences of evidence, not evidence of equality.

The simulation shows sufficiency at most, and this run did not reach it. The dependence rule, fixed before the data and fitted at the end of its grid where an exploratory variable that was zero for most outlets multiplies the cap, reduced to publishing wire copy, so it tested nothing about a resource that varies across outlets, and its margin rested on timing in two cells. The bootstrap share of 0.45 and a within-outlet error equal to the baseline's say that whatever it reproduced lay between outlets, in the inputs it was given. The agents proposed once, from a daily feed, so simulated lags are whole days while observed lags run in hours. Their frame labels agreed too little with the coder to be used. The agent model's training data are not documented by country, so a core-country tilt in its proposals cannot be excluded, and the proposals remain a black box \cite{larooij-2025-large-language_abs}, with the mechanisms kept outside it as rules scored out of sample. A decisive design needs a dependence rule reading a resource that varies across outlets, a timing measure wire items can carry, errors scored on a common set of cells, and agents that adapt over time.

\section{Conclusion}

Wire copy goes with the clock of periphery and semi-periphery foreign news, and the simulation built to separate the mechanisms did not decide between them. Across 18 outlets, six countries and three global issues, one of nine confirmatory tests held. Timing following rose with reliance on AP, AFP and Reuters copy. Tier and local stake showed no effect the intervals could distinguish from none, and periphery outlets followed the core's attention on the issue with the highest local stake. The isomorphism test did not support prestige imitation, with a rule that could not have. The herding test was supported in both wars, with two readings left open. In the simulation the dependence rule met the seed margin, failed the bootstrap condition and, as fitted, reduced to publishing wire copy. What this study secures is narrower than the structural account. Wire copy sets the timing of the periphery's foreign news, and whether dependence or cascade sets its agenda remains to be separated.

The results speak to each of the three lines of research. For intermedia agenda setting, following at the level of timing tracked reliance on wire copy, while following did not differ by tier. For world-system theory, the dependence the theory implies showed up through wire copy, not through tier position as measured. For research on global news flow, the study offers a way to test mechanisms instead of inferring them from correlated agendas: language model agents act as editors, and each mechanism can be switched off while the others stay on.

\section*{Declaration of generative AI use}
This study used generative AI chiefly as a retrieval instrument. A retrieval-augmented system searched a curated library of sources and linked every claim drawn from prior work to a verbatim excerpt verified against its source. Large language models (Anthropic's Claude Fable 5.1 for literature screening; Fable 5.1 and Opus 5.5 for drafting; Fable 5.1 and Opus 5.5 for evidence checking; Fable 5.1 and Opus 5.5 for manuscript review; Opus 5.5 as the coder; Qwen3-14B (Qwen) for the simulation agents; accessed September 2026–October 2026) assisted in screening literature, drafting text from these excerpts, checking whether the evidence supports each claim, and reviewing the manuscript. The authors reviewed, edited and approved all output, confirmed the originality and accuracy of the content and of every reference, and checked the tools' terms of use. The authors take full responsibility for the entire content. Models used to score the corpus are reported in the Method.

\bibliographystyle{ACM-Reference-Format}
\bibliography{references}

\clearpage

\onecolumn

\appendix

\setcounter{table}{0}

\renewcommand{\thetable}{S\arabic{table}}

\section{Design records}\label{app:A}

This appendix holds Tables \ref{tab:S1}, \ref{tab:rules}.

\par\noindent\begin{minipage}{\textwidth}
\centering
\scriptsize
\captionof{table}{Issue Query Terms (Fixed Before Retrieval)}
\label{tab:S1}
\setlength{\tabcolsep}{3pt}
\begin{tabularx}{\textwidth}{>{\raggedright\arraybackslash}p{0.2\textwidth}>{\raggedright\arraybackslash\hsize=1.000\hsize}X}
\toprule
\textbf{Issue} & \textbf{Query} \\
\midrule
War in Ukraine & Ukraine OR Ukrainian OR Kyiv OR Kiev OR Zelensky OR (Russia AND invasion) \\
War in Gaza & Gaza OR Hamas OR (Israel AND (war OR strike OR ceasefire OR hostages)) \\
COP27 / COP28 & COP27 OR COP28 OR (climate AND (summit OR conference) AND (Sharm OR Dubai OR UNFCCC)) \\
\bottomrule
\end{tabularx}
\par\smallskip
\parbox{\textwidth}{\raggedright \emph{Note.} A story belongs to an issue when its title or text matches the query; terms were tested on a pilot week per outlet.}
\end{minipage}\par\bigskip

\par\noindent\begin{minipage}{\textwidth}
\centering
\scriptsize
\captionof{table}{The Three Mechanism Rules Applied to the Agents’ Proposals}
\label{tab:rules}
\setlength{\tabcolsep}{3pt}
\begin{tabularx}{\textwidth}{>{\raggedright\arraybackslash}p{0.12\textwidth}>{\raggedright\arraybackslash\hsize=1.182\hsize}X>{\raggedright\arraybackslash}p{0.08\textwidth}>{\raggedright\arraybackslash\hsize=0.818\hsize}X}
\toprule
\textbf{Mechanism} & \textbf{Rule} & \textbf{Free parameter (grid)} & \textbf{What it does to a day’s proposals} \\
\midrule
Structural dependence & Cap on own foreign stories: B = round(n $\times$ (1 $-$ W) $\times$ (w $\times$ C + 1 $-$ w)), with n the outlet’s mean daily issue stories, W its wire reliance, C its correspondent presence & w (0–1) & Keeps the B highest-priority non-wire proposals; the rest of the foreign news is the day’s wire copy and frames \\
Prestige imitation & Weight of each proposed issue and frame = (1 $-$ v) $\times$ a + v $\times$ s, with a its share in the proposal and s its share among the core outlets that day & v (0–1) & Pulls the outlet’s mix of issues and frames toward the core’s \\
Herding & A new, uncertain story is published, proposed or not, with probability min(1, c $\times$ m / N), with m the outlets that have covered the event and N all outlets; frame = the one most earlier adopters used & c (0.5–5) & Adds stories others have already run, sooner the more have run them \\
\bottomrule
\end{tabularx}
\par\smallskip
\parbox{\textwidth}{\raggedright \emph{Note.} New = within 2 days of the first core story; uncertain = fewer than 3 core stories so far. Baseline: proposals published with no rule. Combined rules apply in the order dependence, imitation, herding. Parameters fitted by grid search on the first validation half and scored on the second.}
\end{minipage}\par\bigskip

\clearpage

\section{Validation and checks on the measures}\label{app:B}

This appendix holds Tables \ref{tab:S3}, \ref{tab:S4}, \ref{tab:S5}, \ref{tab:S6}.

\par\noindent\begin{minipage}{\textwidth}
\centering
\scriptsize
\captionof{table}{Validation of the Coding Under the Model-Stability Protocol}
\label{tab:S3}
\setlength{\tabcolsep}{3pt}
\begin{tabularx}{\textwidth}{>{\raggedright\arraybackslash\hsize=1.556\hsize}X>{\raggedright\arraybackslash\hsize=0.556\hsize}X>{\raggedright\arraybackslash\hsize=1.333\hsize}X>{\raggedright\arraybackslash\hsize=0.667\hsize}X>{\raggedright\arraybackslash\hsize=0.778\hsize}X>{\raggedright\arraybackslash\hsize=1.111\hsize}X}
\toprule
\textbf{Variable} & \textbf{Coded by} & \textbf{Model stability $\alpha$ (3 passes, n = 300)} & \textbf{Prompt-variant $\alpha$} & \textbf{Rule-vs-model $\alpha$ (n)} & \textbf{Meets .80 floor} \\
\midrule
Frame & model & .849 & .799 & — & yes \\
Foreign story & rule, else model & .956 & .406 & 1.000 (1,091) & yes, after one revision \\
Wire credit & rule & .995 & 1.000 & .996 (3,072) & yes \\
National-agency credit & rule & 1.000 & 1.000 & .991 (3,121) & yes \\
Staff foreign byline & rule, else model & .749 & .666 & .944 (1,162) & no — exploratory \\
Agent frame labels vs coder (300 written-out proposals) & agents / model & .232 (agents vs coder, n = 300) & — & — & no — coder frames on written-out proposals \\
Agent recall of sampled outlet headlines (contamination check) & agents & recall at 0.8 similarity = 0.00 (n = 100; mean similarity to the true headline 0.270, to a random one 0.273) & — & — & — \\
\bottomrule
\end{tabularx}
\par\smallskip
\parbox{\textwidth}{\raggedright \emph{Note.} No human coder took part. Model stability is Krippendorff's $\alpha$ (nominal) across three fresh-context passes of the model coder (claude-opus-5-5, prompt version 2) on a random reliability sample of 300 articles stratified by tier; all three passes agreed on 248 of 300 frames. Prompt-variant $\alpha$ compares pass 1 with a reworded prompt. Rule-vs-model $\alpha$ is the agreement between the fixed credit and dateline rules and the model coder on every story the rules decide (n in parentheses, of 3,121). The foreign-story wording was revised once under the rule for a variable below the reliability floor (first wording $\alpha$ = .782). Wire credit excludes the 49 \emph{Washington Post} syndicated copies coded 1 by URL rule from the rule-vs-model count. Agent frame labels: Krippendorff's $\alpha$ between the frame each simulation agent (Qwen/Qwen3-14B) gave its proposal and the model coder's frame for the headline and lead the agent wrote out for it; below the .80 floor, simulated framing is computed from the coder's frames on written-out proposals. Contamination check: the agents were asked to recall sampled outlets' headlines for sampled dates; recall is the share recalled at a token-set similarity of 0.8 or more, with mean similarity to the true headline against a random headline of the same outlet and issue for comparison.}
\end{minipage}\par\bigskip

\par\noindent\begin{minipage}{\textwidth}
\centering
\scriptsize
\captionof{table}{Foreign Stories, Wire Reliance, Correspondent Presence and National-Agency Credits by Outlet}
\label{tab:S4}
\setlength{\tabcolsep}{3pt}
\begin{tabularx}{\textwidth}{>{\raggedright\arraybackslash\hsize=1.522\hsize}X>{\raggedright\arraybackslash\hsize=0.978\hsize}X>{\raggedright\arraybackslash\hsize=0.978\hsize}X>{\raggedright\arraybackslash\hsize=0.761\hsize}X>{\raggedright\arraybackslash\hsize=0.761\hsize}X>{\raggedright\arraybackslash\hsize=0.761\hsize}X>{\raggedright\arraybackslash\hsize=0.761\hsize}X>{\raggedright\arraybackslash\hsize=0.870\hsize}X>{\raggedright\arraybackslash\hsize=1.413\hsize}X>{\raggedright\arraybackslash\hsize=1.196\hsize}X}
\toprule
\textbf{Outlet} & \textbf{Country} & \textbf{Tier} & \textbf{Index score} & \textbf{Coded stories} & \textbf{Foreign stories} & \textbf{Foreign share} & \textbf{Wire reliance} & \textbf{Staff foreign byline (exploratory)} & \textbf{National-agency credit} \\
\midrule
\emph{The Daily Star} & Bangladesh & periphery & 0.34 & 150 & 140 & 0.93 & 0.82 & 0.00 & 0.01 \\
\emph{bdnews24.com} & Bangladesh & periphery & 0.34 & 121 & 120 & 0.99 & 0.56 & 0.00 & 0.00 \\
\emph{New Age} & Bangladesh & periphery & 0.34 & 100 & 97 & 0.97 & 0.00 & 0.00 & 0.00 \\
\emph{The Star (Kenya)} & Kenya & periphery & 0.09 & 150 & 114 & 0.76 & 0.00 & 0.00 & 0.00 \\
\emph{Daily Nation} & Kenya & periphery & 0.09 & 143 & 100 & 0.70 & 0.63 & 0.00 & 0.00 \\
\emph{Tuko} & Kenya & periphery & 0.09 & 133 & 125 & 0.94 & 0.90 & 0.00 & 0.00 \\
\emph{IOL} & South Africa & periphery & 0.41 & 150 & 88 & 0.59 & 0.22 & 0.03 & 0.08 \\
\emph{The Citizen} & South Africa & periphery & 0.41 & 149 & 103 & 0.69 & 0.84 & 0.00 & 0.00 \\
\emph{TimesLIVE} & South Africa & periphery & 0.41 & 150 & 130 & 0.87 & 0.86 & 0.00 & 0.00 \\
\emph{Punch} & Nigeria & semi-periphery & 0.44 & 150 & 98 & 0.65 & 0.60 & 0.00 & 0.03 \\
\emph{Daily Post} & Nigeria & semi-periphery & 0.44 & 150 & 94 & 0.63 & 0.00 & 0.00 & 0.01 \\
\emph{Nigerian Tribune} & Nigeria & semi-periphery & 0.44 & 150 & 49 & 0.33 & 0.00 & 0.00 & 0.00 \\
\emph{Times of India} & India & semi-periphery & 2.83 & 150 & 99 & 0.66 & 0.55 & 0.04 & 0.08 \\
\emph{Hindustan Times} & India & semi-periphery & 2.83 & 149 & 124 & 0.83 & 0.20 & 0.00 & 0.04 \\
\emph{The Hindu} & India & semi-periphery & 2.83 & 150 & 103 & 0.69 & 0.35 & 0.00 & 0.08 \\
\emph{Malay Mail} & Malaysia & semi-periphery & 0.66 & 150 & 139 & 0.93 & 0.94 & 0.00 & 0.03 \\
\emph{New Sarawak Tribune} & Malaysia & semi-periphery & 0.66 & 134 & 78 & 0.58 & 0.28 & 0.03 & 0.58 \\
\emph{Malaysiakini} & Malaysia & semi-periphery & 0.66 & 100 & 39 & 0.39 & 0.08 & 0.00 & 0.54 \\
\emph{The New York Times} & Core & — & — & 144 & 99 & 0.69 & 0.00 & 0.10 & 0.00 \\
\emph{The Washington Post} & Core & — & — & 150 & 101 & 0.67 & 0.52 & 0.19 & 0.00 \\
\emph{The Guardian} & Core & — & — & 150 & 108 & 0.72 & 0.06 & 0.26 & 0.00 \\
BBC News & Core & — & — & 148 & 107 & 0.72 & 0.00 & 0.01 & 0.00 \\
\bottomrule
\end{tabularx}
\par\smallskip
\parbox{\textwidth}{\raggedright \emph{Note.} Shares are over the outlet's coded foreign stories across its kept issue cells. Wire reliance counts credits to AP, AFP or Reuters in the byline, dateline or end note; no text-reuse rule was applied because no agency full texts were available. The 49 \emph{Washington Post} syndicated copies count as wire copy by URL rule. Staff foreign byline is exploratory: its model self-consistency was .749. The index score is Gunaratne's (2001) weighted share of world GNP (55\%) and world exports (45\%) on 2022 World Bank data, in percent-share units.}
\end{minipage}\par\bigskip

\par\noindent\begin{minipage}{\textwidth}
\centering
\scriptsize
\captionof{table}{Checks on the Following Measures by Tier}
\label{tab:S5}
\setlength{\tabcolsep}{3pt}
\begin{tabularx}{\textwidth}{>{\raggedright\arraybackslash\hsize=2.000\hsize}X>{\raggedright\arraybackslash\hsize=0.500\hsize}X>{\raggedright\arraybackslash\hsize=0.500\hsize}X}
\toprule
\textbf{Measure and check} & \textbf{Periphery M (SD)} & \textbf{Semi-periphery M (SD)} \\
\midrule
Salience r, whole period (main) & 0.38 (0.29) & 0.27 (0.31) \\
Salience r, first validation half & 0.41 (0.34) & 0.44 (0.31) \\
Salience r, second validation half & 0.22 (0.25) & 0.23 (0.33) \\
Salience r, core of three (without \emph{The Washington Post}, check) & 0.39 (0.30) & 0.29 (0.31) \\
Salience r, mismatched-issue core series (benchmark) & $-$0.13 (0.13) & $-$0.15 (0.15) \\
Framing similarity, whole period (main) & $-$0.10 (0.09) & $-$0.09 (0.07) \\
Framing similarity, first half & $-$0.08 (0.10) & $-$0.08 (0.10) \\
Framing similarity, second half & $-$0.10 (0.11) & $-$0.08 (0.09) \\
Framing similarity without the 49 headline-only Post copies & $-$0.11 (0.09) & $-$0.10 (0.07) \\
Framing similarity against wire-credited core stories removed (check) & $-$0.11 (0.09) & $-$0.10 (0.08) \\
Framing similarity, non-wire stories only (both samples) & $-$0.09 (0.10) & $-$0.11 (0.09) \\
Framing similarity, mismatched-issue core sample (benchmark) & $-$0.16 (0.09) & $-$0.14 (0.07) \\
Timing: first-story lag (Mdn days), whole period (main) & $-$0.06 (0.13) & $-$0.03 (0.08) \\
Timing: first-story lag, first half & $-$0.08 (0.11) & $-$0.05 (0.09) \\
Timing: first-story lag, second half & $-$0.06 (0.15) & $-$0.01 (0.11) \\
Timing: first-story lag, core first mentions permuted across events (benchmark) & 0.30 (4.68) & $-$0.07 (2.16) \\
Timing: peak cross-correlation lag (weeks), whole period & 0.04 (1.08) & $-$0.15 (0.88) \\
Timing: peak cross-correlation lag, mismatched-issue core series (benchmark) & $-$1.04 (3.10) & 1.08 (3.10) \\
Own-country share of issue-story titles & 0.06 (0.04) & 0.12 (0.08) \\
\bottomrule
\end{tabularx}
\par\smallskip
\parbox{\textwidth}{\raggedright \emph{Note.} Means over kept outlet $\times$ issue cells (periphery 26, semi-periphery 26). Halves: the climate issue splits at June 8, 2023 (COP27 period first, COP28 second); Ukraine at August 13, 2023; Gaza at May 19, 2024 within the war's period from October 7, 2023. The mismatched-issue benchmark pairs each cell with the core series or core sample of a different issue (Ukraine with Gaza, Gaza with COP, COP with Ukraine). The timing benchmark shuffles the core first-mention times across the issue's events (fixed seed). The check with wire-credited core stories removed applies to framing only: census story records carry no credit codes, so salience and timing cannot be recomputed on a wire-free core. The own-country share is computed on story titles, since the census holds no article text.}
\end{minipage}\par\bigskip

\par\noindent\begin{minipage}{\textwidth}
\centering
\tiny
\captionof{table}{Source Check: Media Cloud Series Against GDELT Title-Based Series, by Tier and Cell}
\label{tab:S6}
\setlength{\tabcolsep}{3pt}
\begin{tabularx}{\textwidth}{>{\raggedright\arraybackslash\hsize=0.987\hsize}X>{\raggedright\arraybackslash\hsize=0.538\hsize}X>{\raggedright\arraybackslash\hsize=0.897\hsize}X>{\raggedright\arraybackslash\hsize=1.256\hsize}X>{\raggedright\arraybackslash\hsize=1.077\hsize}X>{\raggedright\arraybackslash\hsize=1.167\hsize}X>{\raggedright\arraybackslash\hsize=1.077\hsize}X}
\toprule
\textbf{Tier / cell} & \textbf{Cells (n) / tier} & \textbf{Media Cloud cells (n) / source} & \textbf{r between the two sources' weekly issue shares, M (SD) / cell r} & \textbf{Salience following, main series, M (SD) / cell} & \textbf{Salience following from GDELT shares, M (SD) / cell} & \textbf{r of main vs GDELT salience following across cells / weeks (n)} \\
\midrule
Periphery & 26 & 18 & 0.68 (0.34) & 0.41 (0.28) & 0.28 (0.28) & 0.70 \\
Semi-periphery & 26 & 26 & 0.87 (0.12) & 0.27 (0.31) & 0.30 (0.24) & 0.18 \\
\emph{Daily Nation} · COP27/COP28 & periphery & Media Cloud & 0.81 & 0.76 & 0.85 & 140 \\
\emph{Daily Nation} · Gaza & periphery & Media Cloud & 0.88 & $-$0.06 & 0.32 & 140 \\
\emph{Daily Nation} · Ukraine & periphery & Media Cloud & 0.62 & 0.22 & 0.38 & 140 \\
\emph{IOL} · COP27/COP28 & periphery & Media Cloud & 0.85 & 0.84 & 0.76 & 127 \\
\emph{IOL} · Gaza & periphery & Media Cloud & 0.91 & 0.43 & 0.02 & 127 \\
\emph{IOL} · Ukraine & periphery & Media Cloud & 0.92 & 0.51 & 0.11 & 127 \\
\emph{New Age} · Gaza & periphery & GDELT (main series) & 1.00 & 0.33 & 0.02 & 130 \\
\emph{New Age} · Ukraine & periphery & GDELT (main series) & 1.00 & 0.27 & 0.25 & 130 \\
\emph{The Citizen} · COP27/COP28 & periphery & Media Cloud & 0.83 & 0.80 & 0.53 & 93 \\
\emph{The Citizen} · Gaza & periphery & Media Cloud & 0.87 & 0.32 & $-$0.02 & 93 \\
\emph{The Citizen} · Ukraine & periphery & Media Cloud & 0.58 & 0.43 & 0.17 & 93 \\
\emph{The Daily Star} · COP27/COP28 & periphery & GDELT (main series) & 1.00 & 0.77 & 0.80 & 149 \\
\emph{The Daily Star} · Gaza & periphery & GDELT (main series) & 1.00 & $-$0.11 & 0.47 & 149 \\
\emph{The Daily Star} · Ukraine & periphery & GDELT (main series) & 1.00 & 0.12 & 0.38 & 149 \\
\emph{The Star} (Kenya) · COP27/COP28 & periphery & Media Cloud & 0.88 & 0.44 & 0.22 & 81 \\
\emph{The Star} (Kenya) · Gaza & periphery & Media Cloud & 0.90 & 0.32 & 0.41 & 81 \\
\emph{The Star} (Kenya) · Ukraine & periphery & Media Cloud & 0.66 & 0.50 & 0.41 & 81 \\
\emph{TimesLIVE} · COP27/COP28 & periphery & Media Cloud & 0.83 & 0.77 & 0.40 & 149 \\
\emph{TimesLIVE} · Gaza & periphery & Media Cloud & 0.88 & 0.42 & 0.18 & 149 \\
\emph{TimesLIVE} · Ukraine & periphery & Media Cloud & 0.94 & 0.54 & 0.54 & 149 \\
\emph{Tuko} · COP27/COP28 & periphery & Media Cloud & $-$0.07 & $-$0.00 & $-$0.18 & 89 \\
\emph{Tuko} · Gaza & periphery & Media Cloud & $-$0.07 & 0.19 & 0.07 & 89 \\
\emph{Tuko} · Ukraine & periphery & Media Cloud & 0.05 & $-$0.01 & $-$0.08 & 89 \\
\emph{bdnews24.com} · COP27/COP28 & periphery & GDELT (main series) & 0.99 & 0.71 & 0.72 & 145 \\
\emph{bdnews24.com} · Gaza & periphery & GDELT (main series) & 0.94 & $-$0.15 & 0.45 & 145 \\
\emph{bdnews24.com} · Ukraine & periphery & GDELT (main series) & 0.91 & 0.41 & 0.41 & 145 \\
\emph{Daily Post} · COP27/COP28 & semi-periphery & Media Cloud & 0.97 & 0.10 & 0.16 & 149 \\
\emph{Daily Post} · Gaza & semi-periphery & Media Cloud & 0.97 & 0.20 & 0.20 & 149 \\
\emph{Daily Post} · Ukraine & semi-periphery & Media Cloud & 0.98 & 0.18 & 0.32 & 149 \\
\emph{Hindustan Times} · COP27/COP28 & semi-periphery & Media Cloud & 0.80 & 0.75 & 0.10 & 130 \\
\emph{Hindustan Times} · Gaza & semi-periphery & Media Cloud & 0.91 & 0.17 & 0.53 & 130 \\
\emph{Hindustan Times} · Ukraine & semi-periphery & Media Cloud & 0.47 & 0.37 & $-$0.04 & 130 \\
\emph{Malay Mail} · COP27/COP28 & semi-periphery & Media Cloud & 0.86 & 0.73 & 0.18 & 147 \\
\emph{Malay Mail} · Gaza & semi-periphery & Media Cloud & 0.91 & 0.60 & $-$0.01 & 147 \\
\emph{Malay Mail} · Ukraine & semi-periphery & Media Cloud & 0.87 & 0.81 & 0.36 & 147 \\
\emph{Malaysiakini} · Gaza & semi-periphery & Media Cloud & 0.84 & 0.13 & 0.20 & 149 \\
\emph{Malaysiakini} · Ukraine & semi-periphery & Media Cloud & 0.85 & $-$0.15 & 0.04 & 149 \\
\emph{New Sarawak Tribune} · COP27/COP28 & semi-periphery & Media Cloud & — & 0.06 & — & 0 \\
\emph{New Sarawak Tribune} · Gaza & semi-periphery & Media Cloud & — & 0.29 & — & 0 \\
\emph{New Sarawak Tribune} · Ukraine & semi-periphery & Media Cloud & — & 0.36 & — & 0 \\
\emph{Nigerian Tribune} · COP27/COP28 & semi-periphery & Media Cloud & 0.61 & 0.33 & 0.62 & 149 \\
\emph{Nigerian Tribune} · Gaza & semi-periphery & Media Cloud & 0.89 & $-$0.22 & 0.27 & 149 \\
\emph{Nigerian Tribune} · Ukraine & semi-periphery & Media Cloud & 0.88 & 0.03 & 0.03 & 149 \\
\emph{Punch} · COP27/COP28 & semi-periphery & Media Cloud & 0.93 & 0.34 & 0.21 & 146 \\
\emph{Punch} · Gaza & semi-periphery & Media Cloud & 0.87 & $-$0.24 & 0.32 & 146 \\
\emph{Punch} · Ukraine & semi-periphery & Media Cloud & 0.93 & 0.22 & 0.15 & 146 \\
\emph{The Hindu} · COP27/COP28 & semi-periphery & Media Cloud & 0.96 & 0.84 & 0.82 & 149 \\
\emph{The Hindu} · Gaza & semi-periphery & Media Cloud & 0.92 & 0.32 & 0.46 & 149 \\
\emph{The Hindu} · Ukraine & semi-periphery & Media Cloud & 0.93 & 0.46 & 0.29 & 149 \\
\emph{Times of India} · COP27/COP28 & semi-periphery & Media Cloud & 0.95 & 0.51 & 0.85 & 149 \\
\emph{Times of India} · Gaza & semi-periphery & Media Cloud & 0.91 & $-$0.28 & 0.46 & 149 \\
\emph{Times of India} · Ukraine & semi-periphery & Media Cloud & 0.93 & 0.23 & 0.34 & 149 \\
\bottomrule
\end{tabularx}
\par\smallskip
\parbox{\textwidth}{\raggedright \emph{Note.} The two tier rows summarize the Media Cloud cells (Bangladesh's three outlets draw their main series from GDELT and are listed with the cells only). For every outlet, GDELT's Global Knowledge Graph host rows give weekly article totals and the number of page titles matching the issue query terms of Table~\ref{tab:S1}, the same measure the Bangladesh outlets use; weeks with fewer than 5 articles are dropped. r between sources is the Pearson correlation of the weekly issue shares from Media Cloud (census stories over the source's story count) and from GDELT titles over the weeks both hold. Salience following from GDELT shares recomputes the event-adjusted salience correlation with both the outlet and the core set measured from GDELT titles (core = mean of the core hosts present; \emph{The Washington Post} included). Columns read per tier row as: cells, Media Cloud cells, mean r between sources, mean main salience following, mean GDELT salience following, correlation of the two across cells; per cell row as: tier, source of the main series, r between sources, main salience following, GDELT salience following, weeks.}
\end{minipage}\par\bigskip

\clearpage

\section{Test output and simulation read-outs}\label{app:C}

This appendix holds Tables \ref{tab:rivals}, \ref{tab:S9}, \ref{tab:S10}, \ref{tab:S11}, \ref{tab:S7}, \ref{tab:S8}.

\par\noindent\begin{minipage}{\textwidth}
\centering
\scriptsize
\captionof{table}{Observational Tests of Prestige Imitation (Isomorphism) and Herding}
\label{tab:rivals}
\setlength{\tabcolsep}{3pt}
\begin{tabularx}{\textwidth}{>{\raggedright\arraybackslash\hsize=1.300\hsize}X>{\raggedright\arraybackslash\hsize=1.200\hsize}X>{\raggedright\arraybackslash\hsize=0.600\hsize}Xr>{\raggedright\arraybackslash}p{0.11\textwidth}r>{\raggedright\arraybackslash\hsize=0.600\hsize}X>{\raggedright\arraybackslash\hsize=1.300\hsize}X}
\toprule
\textbf{Test} & \textbf{Predictor} & \textbf{Level} & \textbf{b} & \textbf{95\% CI} & \textbf{p} & \textbf{n} & \textbf{Decision} \\
\midrule
Isomorphism: framing following on non-wire stories & Intercept (mean framing similarity on non-wire stories) & Framing & $-$0.102 & [$-$0.136, $-$0.068] & < .001 & 52 & not supported (Holm < .001; rule reads a positive intercept as support) \\
Context (no test) & Non-wire minus all-story similarity & Framing & $-$0.006 & — & — & 52 & mean within-cell difference \\
Herding: lag on uncertainty & Core outlets on the event's first day (1–4) & Timing, Ukraine & $-$0.1316 & [$-$0.1726, $-$0.0906] & < .001 & 20,697 events & supported (Holm < .001) \\
Herding: lag on uncertainty & Core outlets on the event's first day (1–4) & Timing, Gaza & $-$0.1640 & [$-$0.1893, $-$0.1387] & < .001 & 22,665 events & supported (Holm < .001) \\
\bottomrule
\end{tabularx}
\par\smallskip
\parbox{\textwidth}{\raggedright \emph{Note.} Isomorphism test: framing similarity recomputed with wire-credited stories removed from both the outlet's and the core sample (bias-corrected as in Table~\ref{tab:following}, so 0 means indistinguishable from the core's non-wire frames); intercept-only mixed model over the 52 kept cells with random intercepts for outlet nested in country; the decision rule reads a positive intercept as support. Herding test: for every GDELT event an outlet covered within one of the two wars (Gaza from October 7, 2023), the negative lag in days from the first core mention to the outlet's first mention (censored at 14 days) regressed on the number of core outlets that mentioned the event on its first core day, the available proxy for the number of core stories that day; random intercepts for outlet nested in country; the decision rule reads a negative coefficient (closer following for less-covered, more uncertain events) as support. Decisions by Holm-adjusted p < .05 across the pair, the herding p being the larger of the two wars.}
\end{minipage}\par\bigskip

\par\noindent\begin{minipage}{\textwidth}
\centering
\scriptsize
\captionof{table}{Full Output: Mixed Models of Following on Tier (H1)}
\label{tab:S9}
\setlength{\tabcolsep}{3pt}
\begin{tabularx}{\textwidth}{>{\raggedright\arraybackslash\hsize=1.481\hsize}X>{\raggedright\arraybackslash\hsize=0.741\hsize}X>{\raggedright\arraybackslash\hsize=0.926\hsize}Xr>{\raggedright\arraybackslash}p{0.1\textwidth}>{\raggedright\arraybackslash\hsize=0.463\hsize}Xr>{\raggedright\arraybackslash\hsize=1.389\hsize}X}
\toprule
\textbf{Model} & \textbf{Predictor} & \textbf{Level} & \textbf{b} & \textbf{95\% CI} & \textbf{p (unadjusted)} & \textbf{Cells (n)} & \textbf{Random effects} \\
\midrule
H1 tier (periphery vs semi-periphery) & Periphery & Salience & 0.103 & [$-$0.148, 0.354] & .421 & 52 & n = 52, var(country) = 0.0142, var(outlet) = 0.0167, residual = 0.0411 \\
H1 check: continuous index score & Index score (per point) & Salience & 0.025 & [$-$0.119, 0.168] & .738 & 52 & n = 52, var(country) = 0.0176, var(outlet) = 0.0165, residual = 0.0412 \\
H1 tier (periphery vs semi-periphery) & Periphery & Framing & $-$0.010 & [$-$0.065, 0.046] & .729 & 52 & n = 52, var(outlet) = 0.0016, residual = 0.0058 \\
H1 tier, main fit before dropping the zero-variance component & Periphery & Framing & $-$0.010 & [$-$0.065, 0.046] & .728 & 52 & n = 52, var(country) = 0.0000, var(outlet) = 0.0016, residual = 0.0058 \\
H1 tier (periphery vs semi-periphery) & Periphery & Framing (without headline-only Post copies) & $-$0.004 & [$-$0.060, 0.051] & .881 & 52 & n = 52, var(outlet) = 0.0015, residual = 0.0060 \\
H1 tier, main fit before dropping the zero-variance component & Periphery & Framing (without headline-only Post copies) & $-$0.004 & [$-$0.060, 0.051] & .881 & 52 & n = 52, var(country) = 0.0000, var(outlet) = 0.0015, residual = 0.0060 \\
H1 check: continuous index score & Index score (per point) & Framing & 0.018 & [$-$0.011, 0.046] & .219 & 52 & n = 52, var(outlet) = 0.0013, residual = 0.0058 \\
H1 check: continuous index, main fit before dropping the zero-variance component & Index score (per point) & Framing & 0.018 & [$-$0.011, 0.046] & .219 & 52 & n = 52, var(country) = 0.0000, var(outlet) = 0.0013, residual = 0.0058 \\
H1 tier (periphery vs semi-periphery) & Periphery & Timing & $-$0.026 & [$-$0.132, 0.081] & .635 & 37 & n = 37, var(country) = 0.0012, var(outlet) = 0.0060, residual = 0.0052 \\
H1 check: continuous index score & Index score (per point) & Timing & 0.007 & [$-$0.049, 0.064] & .794 & 37 & n = 37, var(country) = 0.0014, var(outlet) = 0.0061, residual = 0.0052 \\
\bottomrule
\end{tabularx}
\par\smallskip
\parbox{\textwidth}{\raggedright \emph{Note.} Following per outlet $\times$ issue cell (periphery 26 cells, semi-periphery 26; timing 37 cells with at least 10 matched events). Fixed effects: tier (periphery = 1, semi-periphery = 0) and issue (two dummies, Ukraine and Gaza against COP); random intercepts for outlet nested in country (restricted maximum likelihood). Salience is the event-adjusted correlation, framing the bias-corrected similarity, timing the negative median lag in days, so a positive b means closer following in the periphery. p values are two-sided Wald tests before Holm correction; the Holm-adjusted decisions across the nine confirmatory tests are in Table~\ref{tab:models} and Table~\ref{tab:S11}. A random-effect variance estimated at zero was dropped and the model refitted; both fits are listed. The index check enters Gunaratne's 2022 score (India 2.83, Malaysia 0.66, Nigeria 0.44, South Africa 0.41, Bangladesh 0.34, Kenya 0.09) in place of tier; H1 predicts a negative coefficient.}
\end{minipage}\par\bigskip

\par\noindent\begin{minipage}{\textwidth}
\centering
\scriptsize
\captionof{table}{Full Output: Mixed Models of Following on Wire Reliance (H2) and the First Falsifier}
\label{tab:S10}
\setlength{\tabcolsep}{3pt}
\begin{tabularx}{\textwidth}{>{\raggedright\arraybackslash\hsize=1.316\hsize}X>{\raggedright\arraybackslash\hsize=1.053\hsize}X>{\raggedright\arraybackslash\hsize=0.877\hsize}Xr>{\raggedright\arraybackslash}p{0.1\textwidth}>{\raggedright\arraybackslash\hsize=0.439\hsize}Xr>{\raggedright\arraybackslash\hsize=1.316\hsize}X}
\toprule
\textbf{Model} & \textbf{Predictor} & \textbf{Level} & \textbf{b} & \textbf{95\% CI} & \textbf{p (unadjusted)} & \textbf{Cells (n)} & \textbf{Random effects / read-out} \\
\midrule
H2 wire reliance (confirmatory, no correspondent-presence term) & Wire reliance (0–1) & Salience & $-$0.020 & [$-$0.267, 0.227] & .875 & 52 & n = 52, var(country) = 0.0143, var(outlet) = 0.0188, residual = 0.0412 \\
Tier gap net of wire reliance (first falsifier of the structural account) & Periphery & Salience & 0.108 & [$-$0.153, 0.368] & .418 & 52 & n = 52, var(country) = 0.0143, var(outlet) = 0.0188, residual = 0.0412 \\
First falsifier read-out & Upper CI bound of the net tier gap vs half the H1 estimate & Salience & 0.368 vs 0.052 & — & — & 52 & not met \\
Exploratory: with the correspondent-presence term (outside the Holm family) & Wire reliance & Salience & $-$0.035 & [$-$0.272, 0.203] & .774 & 52 & n = 52, var(country) = 0.0197, var(outlet) = 0.0155, residual = 0.0412 \\
Exploratory: with the correspondent-presence term (outside the Holm family) & Correspondent presence & Salience & $-$1.383 & [$-$3.589, 0.823] & .219 & 52 & n = 52, var(country) = 0.0197, var(outlet) = 0.0155, residual = 0.0412 \\
Exploratory: with the correspondent-presence term (outside the Holm family) & Periphery & Salience & 0.101 & [$-$0.179, 0.381] & .481 & 52 & n = 52, var(country) = 0.0197, var(outlet) = 0.0155, residual = 0.0412 \\
H2 wire reliance (confirmatory, no correspondent-presence term) & Wire reliance (0–1) & Framing & 0.019 & [$-$0.066, 0.104] & .663 & 52 & n = 52, var(outlet) = 0.0019, residual = 0.0057 \\
Tier gap net of wire reliance (first falsifier of the structural account) & Periphery & Framing & $-$0.014 & [$-$0.074, 0.047] & .654 & 52 & n = 52, var(outlet) = 0.0019, residual = 0.0057 \\
H2 wire reliance, main fit before dropping the zero-variance component & Wire reliance (0–1) & Framing & 0.019 & [$-$0.066, 0.104] & .663 & 52 & n = 52, var(country) = 0.0000, var(outlet) = 0.0019, residual = 0.0057 \\
First falsifier read-out & Upper CI bound of the net tier gap vs half the H1 estimate & Framing & 0.047 vs $-$0.005 & — & — & 52 & not met \\
H2 wire reliance (confirmatory, no correspondent-presence term) & Wire reliance (0–1) & Framing (without headline-only Post copies) & 0.025 & [$-$0.061, 0.111] & .571 & 52 & n = 52, var(outlet) = 0.0019, residual = 0.0058 \\
Tier gap net of wire reliance (first falsifier of the structural account) & Periphery & Framing (without headline-only Post copies) & $-$0.010 & [$-$0.070, 0.051] & .758 & 52 & n = 52, var(outlet) = 0.0019, residual = 0.0058 \\
H2 wire reliance, main fit before dropping the zero-variance component & Wire reliance (0–1) & Framing (without headline-only Post copies) & 0.025 & [$-$0.061, 0.111] & .571 & 52 & n = 52, var(country) = 0.0000, var(outlet) = 0.0019, residual = 0.0058 \\
Exploratory: with the correspondent-presence term (outside the Holm family) & Wire reliance & Framing & 0.016 & [$-$0.068, 0.101] & .701 & 52 & n = 52, var(outlet) = 0.0019, residual = 0.0056 \\
Exploratory: with the correspondent-presence term (outside the Holm family) & Correspondent presence & Framing & $-$0.509 & [$-$1.284, 0.266] & .198 & 52 & n = 52, var(outlet) = 0.0019, residual = 0.0056 \\
Exploratory: with the correspondent-presence term (outside the Holm family) & Periphery & Framing & $-$0.017 & [$-$0.077, 0.043] & .588 & 52 & n = 52, var(outlet) = 0.0019, residual = 0.0056 \\
H2 wire reliance (confirmatory, no correspondent-presence term) & Wire reliance (0–1) & Timing & 0.175 & [0.084, 0.267] & < .001 & 37 & n = 37, var(country) = 0.0014, var(outlet) = 0.0018, residual = 0.0051 \\
Tier gap net of wire reliance (first falsifier of the structural account) & Periphery & Timing & $-$0.053 & [$-$0.141, 0.034] & .232 & 37 & n = 37, var(country) = 0.0014, var(outlet) = 0.0018, residual = 0.0051 \\
First falsifier read-out & Upper CI bound of the net tier gap vs half the H1 estimate & Timing & 0.034 vs $-$0.013 & — & — & 37 & not met \\
Exploratory: with the correspondent-presence term (outside the Holm family) & Wire reliance & Timing & 0.174 & [0.081, 0.268] & < .001 & 37 & n = 37, var(country) = 0.0013, var(outlet) = 0.0019, residual = 0.0051 \\
Exploratory: with the correspondent-presence term (outside the Holm family) & Correspondent presence & Timing & $-$0.312 & [$-$1.262, 0.638] & .520 & 37 & n = 37, var(country) = 0.0013, var(outlet) = 0.0019, residual = 0.0051 \\
Exploratory: with the correspondent-presence term (outside the Holm family) & Periphery & Timing & $-$0.053 & [$-$0.140, 0.034] & .231 & 37 & n = 37, var(country) = 0.0013, var(outlet) = 0.0019, residual = 0.0051 \\
\bottomrule
\end{tabularx}
\par\smallskip
\parbox{\textwidth}{\raggedright \emph{Note.} Same cells, levels and random structure as Table~\ref{tab:S9}. Wire reliance is the share of a cell's coded foreign stories that credit AP, AFP or Reuters (credits only; no text-reuse rule), so b is the change in following from a cell with no wire credits to one with wire credits on every foreign story. The confirmatory model carries no correspondent-presence term because correspondent presence fell below the stability floor; the first falsifier therefore reads the tier gap net of wire reliance alone, with correspondent presence inside the remainder. The falsifier read-out compares the upper bound of the net tier gap's 95\% CI with half the H1 estimate from Table~\ref{tab:S9}: the non-wire channels lose support when the bound lies below it and the H1 estimate is positive. The exploratory model adds correspondent presence and stands outside the Holm family. Holm-adjusted decisions are in Table~\ref{tab:models} and Table~\ref{tab:S11}.}
\end{minipage}\par\bigskip

\par\noindent\begin{minipage}{\textwidth}
\centering
\tiny
\captionof{table}{Full Output: Mixed Models of Following on Tier $\times$ Local Stake (H3), the Second Falsifier and Holm Decisions}
\label{tab:S11}
\setlength{\tabcolsep}{3pt}
\begin{tabularx}{\textwidth}{>{\raggedright\arraybackslash\hsize=1.271\hsize}X>{\raggedright\arraybackslash\hsize=0.763\hsize}X>{\raggedright\arraybackslash\hsize=0.847\hsize}Xr>{\raggedright\arraybackslash}p{0.1\textwidth}>{\raggedright\arraybackslash\hsize=0.593\hsize}Xr>{\raggedright\arraybackslash\hsize=1.525\hsize}X}
\toprule
\textbf{Model} & \textbf{Predictor} & \textbf{Level} & \textbf{b} & \textbf{95\% CI} & \textbf{p} & \textbf{Cells (n)} & \textbf{Random effects / decision} \\
\midrule
H1 & Tier (periphery) & Salience & 0.103 & [$-$0.148, 0.354] & 1.000 & 52 & not supported \\
H1 & Tier (periphery) & Framing & $-$0.010 & [$-$0.065, 0.046] & 1.000 & 52 & not supported \\
H1 & Tier (periphery) & Timing & $-$0.026 & [$-$0.132, 0.081] & 1.000 & 37 & not supported \\
H2 & Wire reliance & Salience & $-$0.020 & [$-$0.267, 0.227] & 1.000 & 52 & not supported \\
H2 & Wire reliance & Framing & 0.019 & [$-$0.066, 0.104] & 1.000 & 52 & not supported \\
H2 & Wire reliance & Timing & 0.175 & [0.084, 0.267] & .002 & 37 & supported \\
H3 & Periphery $\times$ stake & Salience & 0.061 & [$-$0.114, 0.236] & 1.000 & 52 & not supported \\
H3 & Periphery $\times$ stake & Framing & 0.019 & [$-$0.032, 0.070] & 1.000 & 52 & not supported \\
H3 & Periphery $\times$ stake & Timing & 0.049 & [$-$0.020, 0.117] & 1.000 & 37 & not supported \\
H3 tier $\times$ stake (linear 0/1/2) & Periphery $\times$ stake & Salience & 0.061 & [$-$0.114, 0.236] & .495 & 52 & n = 52, var(country) = 0.0138, var(outlet) = 0.0095, residual = 0.0669 \\
H3 tier $\times$ stake (linear 0/1/2) & Stake (semi-periphery slope) & Salience & 0.083 & [$-$0.041, 0.207] & .192 & 52 &  \\
H3 tier $\times$ stake (linear 0/1/2) & Periphery (at stake 0) & Salience & 0.042 & [$-$0.260, 0.344] & .785 & 52 &  \\
H3 robustness: two dummy contrasts & Periphery $\times$ stake 1 (Gaza vs Ukraine) & Salience & 0.027 & [$-$0.241, 0.295] & .843 & 52 & n = 52, var(country) = 0.0135, var(outlet) = 0.0174, residual = 0.0420 \\
H3 robustness: two dummy contrasts & Periphery $\times$ stake 2 (COP vs Ukraine) & Salience & 0.133 & [$-$0.146, 0.412] & .349 & 52 & n = 52, var(country) = 0.0135, var(outlet) = 0.0174, residual = 0.0420 \\
H3 robustness: own-country share in place of stake & Periphery $\times$ own-country share & Salience & 2.510 & [$-$0.916, 5.937] & .151 & 52 & n = 52, var(country) = 0.0307, var(outlet) = 0.0187, residual = 0.0594 \\
H3 robustness: own-country share in place of stake & Own-country share (semi-periphery slope) & Salience & 1.023 & [$-$0.291, 2.337] & .127 & 52 &  \\
H3 tier $\times$ stake (linear 0/1/2) & Periphery $\times$ stake & Framing & 0.019 & [$-$0.032, 0.070] & .463 & 52 & n = 52, var(outlet) = 0.0016, residual = 0.0057 \\
H3 tier $\times$ stake (linear 0/1/2) & Stake (semi-periphery slope) & Framing & $-$0.006 & [$-$0.042, 0.030] & .750 & 52 &  \\
H3 tier $\times$ stake (linear 0/1/2) & Periphery (at stake 0) & Framing & $-$0.028 & [$-$0.102, 0.046] & .457 & 52 &  \\
H3 main fit before dropping the zero-variance component & Periphery $\times$ stake & Framing & 0.019 & [$-$0.033, 0.071] & .470 & 52 & n = 52, var(country) = 0.0000, var(outlet) = 0.0016, residual = 0.0057 \\
H3 tier $\times$ stake (linear 0/1/2) & Periphery $\times$ stake & Framing (without headline-only Post copies) & 0.026 & [$-$0.032, 0.083] & .385 & 52 & n = 52, var(country) = 0.0001, var(outlet) = 0.0000, residual = 0.0073 \\
H3 tier $\times$ stake (linear 0/1/2) & Stake (semi-periphery slope) & Framing (without headline-only Post copies) & $-$0.005 & [$-$0.046, 0.035] & .795 & 52 &  \\
H3 tier $\times$ stake (linear 0/1/2) & Periphery (at stake 0) & Framing (without headline-only Post copies) & $-$0.029 & [$-$0.103, 0.046] & .447 & 52 &  \\
H3 robustness: two dummy contrasts & Periphery $\times$ stake 1 (Gaza vs Ukraine) & Framing & 0.052 & [$-$0.060, 0.165] & .364 & 52 & n = 52, var(country) = 0.0002, var(outlet) = 0.0000, residual = 0.0074 \\
H3 robustness: two dummy contrasts & Periphery $\times$ stake 2 (COP vs Ukraine) & Framing & 0.036 & [$-$0.080, 0.152] & .545 & 52 & n = 52, var(country) = 0.0002, var(outlet) = 0.0000, residual = 0.0074 \\
H3 robustness: own-country share in place of stake & Periphery $\times$ own-country share & Framing & $-$0.068 & [$-$0.981, 0.846] & .885 & 52 & n = 52, var(outlet) = 0.0017, residual = 0.0053 \\
H3 robustness: own-country share in place of stake & Own-country share (semi-periphery slope) & Framing & 0.323 & [$-$0.048, 0.695] & .088 & 52 &  \\
H3 tier $\times$ stake (linear 0/1/2) & Periphery $\times$ stake & Timing & 0.049 & [$-$0.020, 0.117] & .165 & 37 & n = 37, var(country) = 0.0012, var(outlet) = 0.0064, residual = 0.0048 \\
H3 tier $\times$ stake (linear 0/1/2) & Stake (semi-periphery slope) & Timing & 0.007 & [$-$0.039, 0.053] & .779 & 37 &  \\
H3 tier $\times$ stake (linear 0/1/2) & Periphery (at stake 0) & Timing & $-$0.058 & [$-$0.173, 0.057] & .322 & 37 &  \\
H3 robustness: two dummy contrasts & Periphery $\times$ stake 1 (Gaza vs Ukraine) & Timing & 0.020 & [$-$0.077, 0.117] & .684 & 37 & n = 37, var(country) = 0.0008, var(outlet) = 0.0068, residual = 0.0049 \\
H3 robustness: two dummy contrasts & Periphery $\times$ stake 2 (COP vs Ukraine) & Timing & 0.135 & [$-$0.023, 0.293] & .093 & 37 & n = 37, var(country) = 0.0008, var(outlet) = 0.0068, residual = 0.0049 \\
H3 robustness: own-country share in place of stake & Periphery $\times$ own-country share & Timing & 0.092 & [$-$1.731, 1.916] & .921 & 37 & n = 37, var(country) = 0.0020, var(outlet) = 0.0054, residual = 0.0059 \\
H3 robustness: own-country share in place of stake & Own-country share (semi-periphery slope) & Timing & $-$0.099 & [$-$0.861, 0.664] & .800 & 37 &  \\
Second falsifier: periphery following at high stake (COP) vs mismatched-issue benchmark & Mean following 0.64 vs benchmark $-$0.20 & Salience & 0.837 & paired t(7) = 9.57 & one-sided < .001, Holm < .001 & 8 & falsifier does not hold (following above benchmark) \\
Second falsifier: periphery following at high stake (COP) vs mismatched-issue benchmark & Mean following $-$0.09 vs benchmark $-$0.12 & Framing & 0.025 & paired t(7) = 1.31 & one-sided .116, Holm .116 & 8 & falsifier holds (following not above benchmark) \\
Second falsifier: periphery following at high stake (COP) vs mismatched-issue benchmark & Mean following 0.04 vs benchmark 0.04 & Timing & — & — & one-sided —, Holm — & 2 & not testable (too few cells) \\
\bottomrule
\end{tabularx}
\par\smallskip
\parbox{\textwidth}{\raggedright \emph{Note.} The first nine rows give the Holm-adjusted decision for each confirmatory test (H1–H3 $\times$ salience, framing, timing; estimates from Tables~\ref{tab:S9} and~\ref{tab:S10} and this table; decision = positive coefficient and Holm-adjusted p < .05). H3 models: stake is coded Ukraine 0, Gaza 1, COP 2 and enters linearly with tier (periphery = 1); issue is dropped because stake and issue coincide; random intercepts for outlet nested in country. A positive periphery $\times$ stake coefficient means the decline in following from low to high stake is smaller among periphery outlets. The dummy-contrast and own-country-share models are robustness checks; the own-country share is the share of an outlet's issue stories whose title names its own country. The second falsifier compares periphery following at high stake (the 8 periphery COP cells; timing 2 with both a value and a permuted benchmark) with the same cells' following against the core series or sample of a different issue (salience: Ukraine core series; framing: Ukraine core sample; timing: core first mentions permuted across events), paired t test, one-sided, Holm-adjusted within the three levels.}
\end{minipage}\par\bigskip

\par\noindent\begin{minipage}{\textwidth}
\centering
\scriptsize
\captionof{table}{Cells With a Simulated Value and Stories Published, by Configuration (Read-Out of the Simulation Run)}
\label{tab:S7}
\setlength{\tabcolsep}{3pt}
\begin{tabularx}{\textwidth}{>{\raggedright\arraybackslash\hsize=1.448\hsize}X>{\raggedright\arraybackslash\hsize=1.034\hsize}X>{\raggedright\arraybackslash\hsize=1.034\hsize}X>{\raggedright\arraybackslash\hsize=1.034\hsize}X>{\raggedright\arraybackslash\hsize=1.034\hsize}X>{\raggedright\arraybackslash\hsize=0.414\hsize}X}
\toprule
\textbf{Configuration} & \textbf{Cells with a salience value (of 52)} & \textbf{Cells with a framing value (of 52)} & \textbf{Cells with a timing value (of 52)} & \textbf{Stories published per seed (M)} & \textbf{Seeds} \\
\midrule
Baseline (no rule) & 52.0 & 52.0 & 52.0 & 72,511 & 20 \\
Dependence & 52.0 & 52.0 & 2.0 & 48,356 & 20 \\
Imitation & 52.0 & 52.0 & 52.0 & 107,519 & 20 \\
Herding & 52.0 & 52.0 & 52.0 & 73,460 & 20 \\
Dependence + imitation & 52.0 & 52.0 & 12.7 & 47,654 & 20 \\
Dependence + herding & 52.0 & 52.0 & 18.0 & 49,122 & 20 \\
Imitation + herding & 52.0 & 52.0 & 52.0 & 108,812 & 20 \\
All three & 52.0 & 52.0 & 28.9 & 48,310 & 20 \\
\bottomrule
\end{tabularx}
\par\smallskip
\parbox{\textwidth}{\raggedright \emph{Note.} Read from the records of run\_20261008\_01 (no new simulation). A cell has a simulated salience value when at least 8 weeks hold at least 5 simulated stories, a framing value when at least 5 published items carry a coder frame, and a timing value when at least 3 published core or event items carry a first-core-mention lag; wire items carry no lag. The standardized errors in Table~\ref{tab:sim} are means over the cells that hold a value, so the timing error of a dependence configuration rests on the cells listed here. Means over the 20 seeds, with rules applied in the order dependence, imitation, herding.}
\end{minipage}\par\bigskip

\par\noindent\begin{minipage}{\textwidth}
\centering
\scriptsize
\captionof{table}{Effective Daily Cap on Own Stories Under the Fitted Dependence Rule, by Outlet}
\label{tab:S8}
\setlength{\tabcolsep}{3pt}
\begin{tabularx}{\textwidth}{>{\raggedright\arraybackslash\hsize=1.143\hsize}X>{\raggedright\arraybackslash\hsize=0.667\hsize}X>{\raggedright\arraybackslash\hsize=0.762\hsize}X>{\raggedright\arraybackslash\hsize=1.333\hsize}X>{\raggedright\arraybackslash\hsize=1.143\hsize}X>{\raggedright\arraybackslash\hsize=0.952\hsize}X}
\toprule
\textbf{Outlet} & \textbf{Wire reliance W} & \textbf{Correspondent presence C} & \textbf{Cap B at the fitted w = 1.00 (Ukraine / Gaza / COP)} & \textbf{Cap B at w = 0 (Ukraine / Gaza / COP)} & \textbf{Keeps no own story at the fitted w} \\
\midrule
\emph{bdnews24.com} & 0.57 & 0.00 & 0 / 0 / 0 & 1 / 1 / 0 & yes \\
\emph{The Citizen} & 0.84 & 0.00 & 0 / 0 / 0 & 0 / 1 / 0 & yes \\
\emph{Daily Post} & 0.00 & 0.00 & 0 / 0 / 0 & 2 / 1 / 0 & yes \\
\emph{Hindustan Times} & 0.20 & 0.00 & 0 / 0 / 0 & 11 / 24 / 0 & yes \\
\emph{IOL} & 0.22 & 0.03 & 0 / 0 / 0 & 5 / 4 / 0 & yes \\
\emph{Malay Mail} & 0.94 & 0.00 & 0 / 0 / 0 & 1 / 1 / 0 & yes \\
\emph{Malaysiakini} & 0.11 & 0.00 & 0 / 0 / 0 & 0 / 0 / 0 & yes \\
\emph{Daily Nation} & 0.59 & 0.00 & 0 / 0 / 0 & 1 / 1 / 0 & yes \\
\emph{New Age} & 0.00 & 0.00 & 0 / 0 / 0 & 0 / 1 / 0 & yes \\
\emph{New Sarawak Tribune} & 0.27 & 0.04 & 0 / 0 / 0 & 0 / 1 / 0 & yes \\
\emph{Punch} & 0.52 & 0.00 & 0 / 0 / 0 & 2 / 1 / 0 & yes \\
\emph{The Star (Kenya)} & 0.00 & 0.00 & 0 / 0 / 0 & 6 / 4 / 0 & yes \\
\emph{The Daily Star} & 0.82 & 0.00 & 0 / 0 / 0 & 0 / 1 / 0 & yes \\
\emph{The Hindu} & 0.35 & 0.00 & 0 / 0 / 0 & 9 / 7 / 0 & yes \\
\emph{TimesLIVE} & 0.86 & 0.00 & 0 / 0 / 0 & 0 / 0 / 0 & yes \\
\emph{Times of India} & 0.55 & 0.05 & 1 / 1 / 0 & 13 / 16 / 0 & no \\
\emph{Nigerian Tribune} & 0.00 & 0.00 & 0 / 0 / 0 & 1 / 1 / 0 & yes \\
\emph{Tuko} & 0.92 & 0.00 & 0 / 0 / 0 & 0 / 0 / 0 & yes \\
\bottomrule
\end{tabularx}
\par\smallskip
\parbox{\textwidth}{\raggedright \emph{Note.} B = round(n $\times$ (1 $-$ W) $\times$ (w $\times$ C + 1 $-$ w)), with n the outlet's mean daily count of issue stories in the fitting half. At the fitted w = 1.00 the cap is n $\times$ (1 $-$ W) $\times$ C, so an outlet with no staff foreign byline (C = 0.00) keeps no own proposal and publishes wire copy only. 17 of 18 outlets keep no own story at the fitted w. Correspondent presence is exploratory; it was fixed as the rule's C before the data.}
\end{minipage}\par\bigskip

\end{document}